\documentclass[11pt]{article}
\usepackage[T1]{fontenc}
\usepackage[margin=1in]{geometry}
\usepackage{amsmath,amssymb,amsthm,mathtools}
\usepackage{bm,bbm}
\numberwithin{equation}{section}
\usepackage{microtype}
\usepackage{array,booktabs}
\usepackage{tikz}
\usepackage[numbers,sort&compress]{natbib}
\usepackage[colorlinks=true,linkcolor=blue,citecolor=blue,urlcolor=blue,pagebackref=false]{hyperref}
\usepackage[shortlabels]{enumitem}
\usepackage{authblk}

\newtheorem{theorem}{Theorem}
\newtheorem{lemma}[theorem]{Lemma}
\newtheorem{corollary}[theorem]{Corollary}
\theoremstyle{remark}

\DeclareMathOperator{\Tr}{Tr}
\DeclareMathOperator{\tr}{tr}
\DeclareMathOperator{\diam}{diam}
\DeclareMathOperator{\dist}{dist}
\DeclareMathOperator{\supp}{supp}
\def \id{\mathbbm{1}}
\def \ind{\mathbf{1}}
\newcommand{\E}{\mathbb E}
\newcommand{\norm}[1]{\left\lVert #1\right\rVert}
\newcommand{\ket}[1]{\ensuremath{\left|{#1}\right\rangle}}
\newcommand{\bra}[1]{\ensuremath{\left\langle{#1}\right|}}
\newcommand{\ketbra}[1]{\ensuremath{\ket{#1}\hspace{-2pt}\bra{#1}}}
\newcommand{\lrp}[1]{\left( {#1} \right)}
\newcommand{\lrb}[1]{\left[ {#1} \right]}
\def \s{\mathbf{s}}

\def \C{\mathsf{C}}
\def \D{\mathsf{D}}
\def \F{\mathsf{F}}
\def \K{\mathsf{K}}
\title{High-temperature Gibbs states are uniformly Markovian and trivial}
\author[1]{Andrew D. Kim\textsuperscript{*}}
\author[1]{Tai Hsuan Yang\textsuperscript{*}}
\author[1,2]{Jong Yeon Lee}
\affil[1]{Department of Physics and Institute for Condensed Matter Theory, University of Illinois Urbana-Champaign, Urbana, Illinois 61801, USA}
\affil[2]{Korea Institute for Advanced Study, Seoul 02455, South Korea}

\date{}

\begin{document}
\maketitle

\begingroup
\renewcommand{\thefootnote}{*}
\footnotetext{These authors contributed equally.}
\endgroup

\vspace{-30pt}

\begin{abstract}
We prove that, above a constant temperature threshold, Gibbs states of local noncommuting qudit Hamiltonians exhibit exponential decay of mutual information and conditional mutual information (CMI).
Specifically, we prove exponential decay of the CMI for arbitrary spatial tripartitions $A \sqcup B \sqcup C$ which need not comprise the entire system, thereby establishing a \textit{uniform Markov property} for high-temperature Gibbs states.
A central obstruction to proving such bounds is that conventional cluster expansions for the logarithm of reduced Gibbs states can fail to converge due to noncommutativity. 
Our main insight is to construct a classical representation of the Gibbs state using an overcomplete basis. 
Although such representations generally involve negative quasiprobability weights, we show that a decomposition based on projective 2-designs admits a convergent cluster expansion with strictly positive weights at sufficiently high temperatures, allowing us to circumvent the obstruction.
The separability of high-temperature Gibbs states is a natural byproduct of this decomposition.
Finally, in the context of mixed-state phases of matter, our results imply that high-temperature Gibbs states belong to the same mixed-state phase as the maximally mixed state.
\end{abstract}

\begingroup
\small
\setlength{\parskip}{0pt}
\renewcommand{\baselinestretch}{0.92}\selectfont
\tableofcontents
\endgroup


\section{Introduction}\label{sec:introduction}
The quantum Gibbs state of a Hamiltonian $H$
\begin{align}
    \rho_\beta := \frac{e^{-\beta H}}{\Tr\left\{e^{-\beta H}\right\}} \label{eq:Gibbs_state_defn}
\end{align}
describes thermal equilibrium at inverse temperature $\beta$, and therefore understanding its general structure is a central problem in quantum many-body physics. 
Substantial progress has been made in establishing its structural properties, including clustering of correlations and mutual information~\cite{Araki1969,KlieschEtAl2014,FrohlichUeltschi2015,BluhmCapelPerezHernandez2022,BluhmCapelPerezHernandez2025StrongDecay}, mutual-information area laws~\cite{Cirac_arealaw_2008,KuwaharaAlhambraAnshu2021AreaLaw,YousefiRezakhani2026AreaLaw}, Markov properties~\cite{BrownPoulin2012,KatoBrandao2019,Kuwahara2025,KatoKuwahara2025,Scalet2025,BakshiLiuMoitraTang2025,Chen2025,rosaruiz2026staticfeaturesmixingshort,Yang2026LocalMarkov,bergamaschi2025structuraltheoryquantummetastability}, sudden death of entanglement~\cite{AndersWinter2007Thermal,Anders2008Thermal,BakshiLiuMoitraTang2024,RamkumarEtAl2025Gaussian,PuttermanZlokapaCotler2026Classical,Kiani2026SharpEntanglement}, and the sample and computational complexity of learning its Hamiltonian~\cite{AnshuEtAl2020Learning,HaahKothariTang2024Learning,BakshiEtAl2023Learning,Narayanan2024Learning,ChenAnshuNguyen2025Learning,WangYeChen2026MetastableLearning,Bergamaschi2026Stationarity,LewisTangWright2026Learning,DongLeng2026ScoreMatching,yang2026nearoptimallearningquantummanybody}.

Because physical systems exist at finite temperature, classifying quantum Gibbs states using a suitable framework for mixed-state phases of matter is a broad and important direction.
In such frameworks, one requires various information-theoretic properties, either structural or dynamical~\cite{yang2025topologicalmixedstatesphases,Ma2025Gibbs,Sang2024,sang2025mixedstatephaseslocalreversibility}. 
Two such properties are the decay of mutual information (MI)
\begin{align}
    I(A:C)_\rho \leq f_{\mathrm{MI}}(|A|,|C|) \cdot \delta_{\mathrm{MI}}\left(\dist(A,C) \right) \label{eq:MI_decay_defn}
\end{align}
and the decay of conditional mutual information (CMI)
\begin{align}
    I(A:C | B)_{\rho} \leq f_{\mathrm{CMI}}(|A|,|C|) \cdot \delta_{\mathrm{CMI}}\left(\dist(A,C)\right) \label{eq:CMI_decay_defn}
\end{align}
for a state $\rho$ on a set of qudits $\Lambda$.
In Eqs.~\eqref{eq:MI_decay_defn} and \eqref{eq:CMI_decay_defn}, $\delta_{\mathrm{MI}}$ and $\delta_{\mathrm{CMI}}$ are decaying functions, and the functions $f_{\mathrm{MI}}$ and $f_{\mathrm{CMI}}$ grow at most polynomially (and may actually depend only on $|\partial A|, |\partial C|$).
For Gibbs states $\rho = \rho_\beta$, rigorous proofs of these properties have been established in 1D, for commuting Hamiltonians, or for specific partition families.
However, they have remained unproven at the level of generality required to classify finite-temperature quantum phases; see Sec. \ref{sec:discussion} for more discussion.

Table~\ref{tab:information-decay-comparison} compares five representative results with the bounds established here.
The distinction concerns both the size dependence of the prefactor and the allowed partitions: a CMI bound for a full partition does not by itself control the CMI after part of the conditioning region is traced out.
Our high-temperature bounds combine exponential distance decay with a decaying rate independent of the temperature and with a prefactor linear in the smaller interaction boundary uniformly over arbitrary disjoint regions.

\begin{table}[!htbp]
\centering
\footnotesize
\setlength{\tabcolsep}{4pt}
\renewcommand{\arraystretch}{1.18}
\newlength{\comparisonReferenceWidth}
\newlength{\comparisonSettingWidth}
\newlength{\comparisonBoundWidth}
\newlength{\comparisonPartitionWidth}
\setlength{\comparisonReferenceWidth}{0.19\textwidth}
\setlength{\comparisonSettingWidth}{0.25\textwidth}
\setlength{\comparisonBoundWidth}{0.30\textwidth}
\setlength{\comparisonPartitionWidth}{\dimexpr\textwidth-\comparisonReferenceWidth-\comparisonSettingWidth-\comparisonBoundWidth-6\tabcolsep\relax}
\begin{tabular}{@{}>{\raggedright\arraybackslash}p{\comparisonReferenceWidth}
                  >{\raggedright\arraybackslash}p{\comparisonSettingWidth}
                  >{\raggedright\arraybackslash}p{\comparisonBoundWidth}
                  >{\raggedright\arraybackslash}p{\comparisonPartitionWidth}@{}}
\toprule
\textbf{Reference} & \textbf{Hamiltonian and regime} & \textbf{Information bound} & \textbf{Regions / partition} \\
\midrule
Kato--Brand\~ao~\cite{KatoBrandao2019}
& 1D, finite range; all temperatures
& CMI: $C_\beta e^{-c_\beta\sqrt{r}}$
& Consecutive regions $A$--$B$--$C$ in a chain \\
\addlinespace
Bluhm--Capel--P\'erez-Hern\'andez~\cite{BluhmCapelPerezHernandez2025StrongDecay}
& Finite range, high temperature, commuting
& MI: $\log(1+u_\beta e^{u_\beta})e^{-c_\beta r}$\newline
  $u_\beta=C_\beta\min\{|\partial A|,|\partial C|\}$
& Arbitrary disjoint $A,C\subseteq\Lambda$ \\
\addlinespace
Chen--Rouz\'e~\cite{Chen2025}
& Finite range, all temperatures
& CMI: $C_\beta |A||C| e^{C_\beta\min\{|A|,|C|\}-c_\beta r}$
& Full partition; $B$ shields $A$ from $C$ \\
\addlinespace
Yang~\cite{Yang2026LocalMarkov}
& Finite range, all temperatures
& CMI: $C_\beta e^{C_\beta\min\{|\partial A|,|\partial C|\}-c_\beta r}$
& Full partition; $B$ shields $A$ from $C$ \\
\addlinespace
Bakshi--Liu--Moitra--Tang~\cite{BakshiLiuMoitraTang2025}
& Finite range; high temperature
& CMI: $C_\beta |A||C|e^{-c_\beta r}$
& Full partition; $B$ shields $A$ from $C$ \\
\midrule
\textbf{This work}
& Finite range; high temperature
& \textbf{MI and CMI:}\newline
  $C_\beta\min\{|\partial A|,|\partial C|\}e^{-r/R}$
& Arbitrary disjoint $A,B,C\subseteq\Lambda$ \\
\bottomrule
\end{tabular}
\caption{Selected bounds on mutual information and conditional mutual information of quantum Gibbs states.
Here $r=\dist(A,C)$ and $R$ is the diameter of local Hamiltonian terms (and is thus independent of $\beta$).
A full partition means $\Lambda=A\sqcup B\sqcup C$.
The constants $C_\beta,c_\beta>0$ may differ between rows and depend on temperature and fixed local parameters. 
We use $\partial A$ for the boundary of $A$ in each result. The last row summarizes Theorems~\ref{thm:MI_decay} and~\ref{thm:CMI_decay} at their respective temperature thresholds.\\
\noindent 
}
\label{tab:information-decay-comparison}
\end{table}

In the above definitions, the functions $\delta$ are at best exponentially decaying, but this cannot be expected for all temperatures or Hamiltonians. 
At the finite-temperature critical point of the two-dimensional Ising model, spin-spin correlations decay algebraically~\cite{Wu1966IsingCorrelations}, ruling out exponential decay of MI.
Moreover, for some tripartitions $A \sqcup B \sqcup C \subseteq \Lambda$, the CMI can retain a nonzero topological contribution even when local correlations are short-ranged~\cite{LevinWen2006,KitaevPreskill_2006,yang2025topologicalmixedstatesphases}; such contributions occur in topological ground states and can survive at low temperature in models with thermally stable topological entropy~\cite{CastelnovoChamon2008ThermalTopology,AlickiEtAl2008ThermalStability,BalasubramanianDavydovaLin2026Memory}.

Consequently, a logical first step towards classifying finite-temperature quantum phases is the study of Gibbs states of local Hamiltonians at \textit{high temperature}.
In this setting, one may still expect the exponential decay of MI and CMI and thus that $\rho_\beta$ lies in a `trivial' mixed-state phase.
Indeed, for local Hamiltonians with \textit{commuting} terms\footnote{Actually, for these results (summarized in row 2 of Table~\ref{tab:information-decay-comparison}) to hold, a stronger assumption that interaction terms in $H$ belong to a commutative algebra which is closed under partial traces is required.}, high-temperature Gibbs states are known to exhibit MI decay~\cite{BluhmCapelPerezHernandez2025StrongDecay}.
It has further been shown that CMI decay persists in the presence of weak local channels~\cite{ZhangGopalakrishnan2026HiddenMarkov}, indicating that they belong to a stable trivial phase.
However, the cluster expansion technique utilized there cannot be applied in the noncommutative setting, as the corresponding operator-valued cluster expansion need not converge uniformly in system size~\cite{KuwaharaKatoBrandao2020,KuwaharaKatoBrandao2025Erratum,KatoKuwahara2025}.

Furthermore, for full tripartitions in which $B$ shields $A$ from $C$, commuting local Hamiltonians admit a block decomposition of the conditioning system that makes quantum conditional independence explicit. 
This structure is generally absent for noncommuting Hamiltonians and need not persist when part of the conditioning region is traced out, complicating the analysis of CMI for arbitrary reduced states.
This may seem to suggest that, even at high temperature, more sophisticated techniques are needed to rigorously classify mixed-state phases of Gibbs states of noncommuting Hamiltonians.

On the other hand, recent results have proven the separability of (and thus absence of entanglement in) high-temperature Gibbs states. 
That is, for local Hamiltonians $H$ on $n$ qudits and $\beta$ below a constant inverse temperature threshold, the Gibbs state $\rho_\beta$ admits a separable decomposition~\cite{BakshiLiuMoitraTang2024}
\begin{align}
    \rho_\beta = \sum_{\s} \pi(\s) \, \mathsf{A}_{s_1} \otimes \cdots \otimes \mathsf{A}_{s_n} \label{eq:Gibbs_separability_BLMT24}
\end{align}
where $\mathsf{A}_{s_i}$ is a local projector for qudit $i$ and $\pi(\s) > 0$ is a probability distribution over classical projector labels $\s=(s_1,\ldots,s_n)$.
This result suggests that, in principle, establishing the decay of MI and CMI for high-temperature Gibbs states of noncommuting Hamiltonians can be done by obtaining a more refined structural understanding of the classical probability distribution $\pi(\s)$.
That is, a distribution $\pi(\s)$ which can be straightforwardly marginalized can be used to bound $I(A:C)_{\rho_\beta}$ and $I(A:C|B)_{\rho_\beta}$.

In this work, we will follow such an approach to establish the exponential decay of MI and CMI (for arbitrary tripartitions) in high-temperature Gibbs states of local noncommuting Hamiltonians.
Our main technical contribution is the use of a local projective 2-design to decompose $\rho_\beta$ into an expectation over products of local projectors with scalar Gibbs weights, each corresponding to a classical label $\s$.
We perform a cluster expansion on the Gibbs weight for each label $\s$, and then we show that each cluster expansion converges below a constant value of $\beta$. 
As a consequence, each weight is positive, and thus the Gibbs state is separable at high temperature.

With the aim of proving high-temperature MI and CMI decay, this allows us to express the resulting distribution $\pi(\s)$ via the cluster-expanded weights, and we can directly obtain controlled and structured expressions for its marginals by averaging out local classical labels.
Then, for any $L \subseteq \Lambda$, we can express the reduced Gibbs state $\rho_{\beta,L}$ using the marginals of $\pi(\s)$, which allows us to handle subsystem entropy contributions of the form $\log(\rho_{\beta,L})$ and thus prove exponentially decaying upper bounds for the MI and CMI.
Notably, our approach avoids divergence issues from the logarithm of reduced Gibbs states.

\subsection{Relation to other works}

\subsubsection*{Decay of correlations and (conditional) mutual information}\label{subsec:review_MI_decay}
Standard cluster expansions express the logarithm of the partition function---and consequently connected correlation functions---as a series over connected subsets of interactions in which disconnected contributions cancel identically.
At sufficiently high temperature, these series converge absolutely with cluster weights decaying exponentially in their size; since any cluster which couples two distant regions must span their separation, this directly yields exponential decay of correlations~\cite{KP,Ueltschi,KlieschEtAl2014,FrohlichUeltschi2015}.
Extending this reasoning to decay of MI and CMI is not straightforward, because their entropy combinations involve logarithms of reduced Gibbs states.
A generalized cluster expansion for these effective subsystem Hamiltonians was proposed~\cite{KuwaharaKatoBrandao2020}, but subsequent work revealed that noncommutativity leads to divergences in the resulting expansions~\cite{KuwaharaKatoBrandao2025Erratum,KatoKuwahara2025}.
Establishing the exponential decay of MI and CMI (for generic partitions) in high-temperature Gibbs states of local noncommuting Hamiltonians has therefore remained open prior to this work.

\subsubsection*{Decay of conditional mutual information from dynamics}\label{subsec:review_CMI_decay}
Recent developments in quantum Gibbs sampling provide a dynamical perspective on CMI decay.
Following the first construction of an efficiently implementable Lindbladian satisfying exact detailed balance with respect to Gibbs states of noncommuting Hamiltonians~\cite{ChenKastoryanoGilyen2023,ChenKastoryanoBrandaoGilyen2023,ChenEtAl2025ThermalSimulation}, subsequent work extended this framework to discrete jump operators and discrete-time quantum Glauber and Metropolis dynamics~\cite{DingLiLin2025,GilyenChenDoriguelloKastoryano2024,JiangIrani2024Metropolis}.
The cost of preparing the Gibbs state depends on the mixing time of these dynamics.
In addition to polynomial and logarithmic mixing-time bounds for local Hamiltonians at high temperature~\cite{RouzeFrancaAlhambra2026Thermalization,RouzeFrancaAlhambra2026Optimal,BakshiLiuMoitraTang2025}, mixing-time estimates in various other regimes have also been established~\cite{TongZhan2025Fermions,SmidEtAl2025FermiHubbard,SmidEtAl2025WeakMixing,BergamaschiChen2025SpinChains,Kim2026RapidMixingChains,RouzeFranca2026DobrushinShlosman,ChenDing2026SpectralMixing,CapelPerezGarciaScandi2026MLSI,Bergamaschi2026AllToAll,CaiZhanZlokapa2026SYK,RamkumarSoleimanifar2025Mixing}.

It is therefore natural to ask how dynamical properties constrain the static correlations of Gibbs states. 
Indeed, several recent works have established such connections. 
Under the assumption of uniform rapid mixing of a quantum Gibbs sampler (or uniform clustering) for all subsystem Hamiltonians, Ref.~\cite{KatoKuwahara2025} reduced CMI decay to the quasi-locality of quantum belief-propagation channels, which yields stretched-exponential CMI decay for full tripartitions $A\sqcup B\sqcup C=\Lambda$ in which $B$ shields $A$ from $C$. This was subsequently sharpened to exponential decay at sufficiently high temperature~\cite{BakshiLiuMoitraTang2025}.
More recent work has studied both MI and CMI in steady states of more general frustration-free Lindbladians under rapid mixing assumptions~\cite{rosaruiz2026staticfeaturesmixingshort}.

However, all aforementioned CMI bounds (from Gibbs sampling dynamics or otherwise) apply to shielding full tripartitions $A \sqcup B \sqcup C = \Lambda$.
In contrast, the CMI bound of our Theorem~\ref{thm:CMI_decay} (proven without constructing or analyzing any Gibbs sampling dynamics) holds for arbitrary tripartitions $A \sqcup B \sqcup C \subseteq \Lambda$ with all remaining sites $D$ traced out: a \textit{uniform Markov property}~\cite{Brandao_Kastoryano_2018} for high-temperature Gibbs states of generic local Hamiltonians.
Whether rapid mixing of a local Lindbladian alone implies this stronger, boundary-law CMI decay remains an interesting open question.

\subsubsection*{Separability of high-temperature Gibbs states}
Separability of Gibbs states of local Hamiltonians at high temperature was first proven algorithmically via an adaptive pinning procedure~\cite{BakshiLiuMoitraTang2024}.
Subsequent work has sharpened the temperature threshold~\cite{PuttermanZlokapaCotler2026Classical,Kiani2026SharpEntanglement}, extended the approach to long-range Hamiltonians~\cite{PuttermanZlokapaCotler2026Classical}, and generalized it to fermionic systems~\cite{RamkumarEtAl2025Gaussian}.
Along the lines of recent work~\cite{Kiani2026SharpEntanglement}, to obtain separability, we first decompose the Gibbs state $\rho_\beta$ using a particular choice of operator basis and then directly prove that all of its coefficients are positive for $\beta$ below a constant threshold.   

\begin{figure}[!t]
    \centering
    \includegraphics[width=0.98\linewidth]{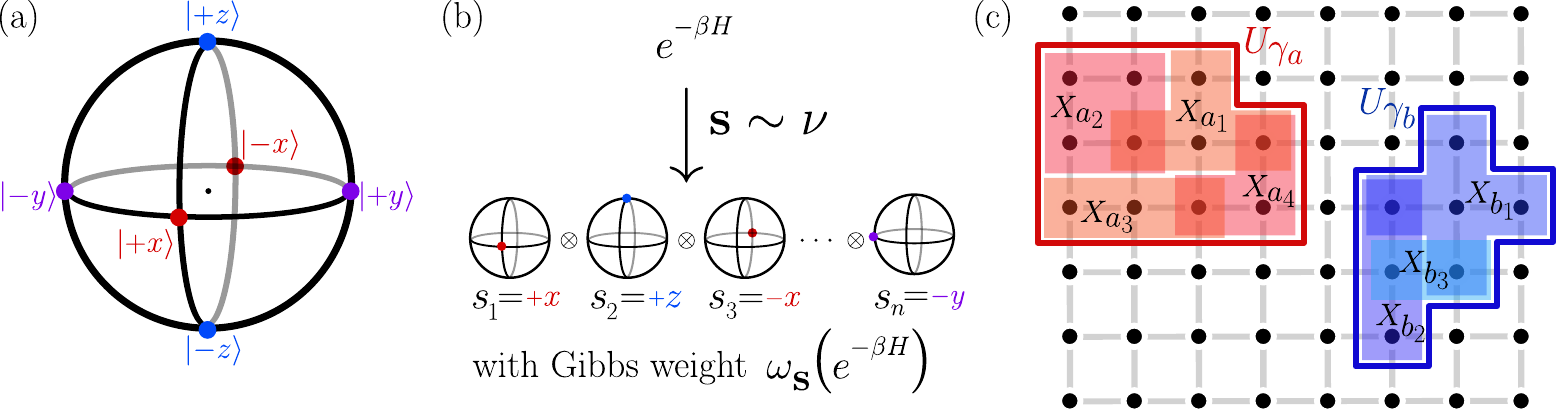}
    \caption{
    (a) A simple example of an on-site projective 2-design for qubits on the Bloch sphere; assign a label to the six Pauli eigenstate projectors, and give each of them probability $1/6$.
    (b) The Gibbs weights $\omega_\s(e^{-\beta H})$ of the 2-design operator decomposition can be thought of as measured coefficients of the Gibbs operator $e^{-\beta H}$ obtained from a classical shadows snapshot.
    (c) We use a cluster expansion to control the scalar Gibbs weights; pictured is an example of two compatible polymers $\gamma_a, \gamma_b$ in the cluster expansion of $e^{-\beta H}$.
    $\{ X_{a_i} \}$ are the supports of Hamiltonian terms, and $U_{\gamma_a}$ is the total support of $\gamma_a$.
    The weight functional $\omega_\s$ factorizes over compatible polymers, and the resulting factor for each depends only on the local labels for its support.
    }
    \label{fig:2-design}
\end{figure}

The 2-design decomposition we use defines a natural single-site measurement (for qubits, a measurement along a randomly chosen Pauli axis; see Figure~\ref{fig:2-design}a), and the dual operators used to extract Gibbs state coefficients are precisely the classical shadow ``snapshots'' for learning quantum states~\cite{Huang2020Shadows,Innocenti2023shadowtomography}. 
We use the same two sets of operators as classical shadows, but in opposite roles.
The classical shadows formalism represents a state as an average of snapshots weighted by measurement probabilities; this works for any state, but the snapshots themselves are not physical states.
Instead, we represent $\rho_\beta$ as an average of physical product states, weighted by coefficients obtained by tracing the Gibbs operator $e^{-\beta H}$ against the snapshots, as depicted in Figure~\ref{fig:2-design}b.
We guarantee positivity of each coefficient at high temperature using a cluster expansion (Figure~\ref{fig:2-design}c), which implies separability.

We remark that our temperature threshold for separability is not tight; our 2-design decomposition incurs an extra factor of $(2q-1)^{-k}$ (where $q$ is the qudit dimension and $k$ is the Hamiltonian locality) compared to sharper thresholds obtained in recent work~\cite{PuttermanZlokapaCotler2026Classical,Kiani2026SharpEntanglement}.
This factor is inherent to the 2-design decomposition, so there does not appear to be room for qualitative improvement of our temperature thresholds.
Also, we do not explicitly construct a classical algorithm to sample from high-temperature Gibbs states (although we believe such an algorithm can be straightforwardly obtained using the structure of $\pi(\s)$ and its marginals).

\section{Results}\label{sec:results}
\subsection{Setting and definitions}\label{subsec:setting}
Let $\Lambda$ be a finite subset of a finite-dimensional lattice $\mathbb{L}$ with distance metric $\dist(\cdot,\cdot)$ and polynomial volume growth.
We define the lattice sum
\begin{align}
 \mathsf{S}_t := \sup_{x \in \mathbb{L}} \sum_{y \in \mathbb{L}} e^{-\dist(x,y)/t} \qquad (t>0) \label{eq:lattice_kernel}
\end{align}
for use in the statement of our main results.
The sum $\mathsf{S}_t$ is convergent for finite-dimensional lattices $\mathbb{L}$ (see Appendix~\ref{app:lattice_kernel} for a short derivation).

On each site $x \in \Lambda$, we define a local Hilbert space $\mathcal{H}_x$ of finite dimension $d_x$ which is at most $q$:
\begin{align}
 \mathcal{H} = \bigotimes_{x\in\Lambda} \mathcal{H}_x,
 \qquad d_x:= \dim(\mathcal{H}_x) \leq q. \label{eq:Hilbert_space_defn}
\end{align}
We consider Hamiltonians 
\begin{align}
    H = \sum_{X \in \mathcal{E}} h_X \label{eq:Hamiltonian_defn}
\end{align}
acting on $\mathcal{H}$ which are $k$-local, have range $R$, and have bounded interaction strength per site. 
Explicitly, $\mathcal{E}$ is a finite set of (nonempty) interaction supports $X \subseteq \Lambda$, and each term $h_X = h_X^\dag$ is a Hermitian operator with $\supp(h_X) = X$.
We make no assumption about the commutation relation between pairs of terms in $H$.
The locality $k$ and range $R$ are positive integers such that, for any $X \in \mathcal{E}$,
\begin{align}
    |X| \leq k, \qquad \diam(X) \leq R.
\end{align}
The total interaction strength experienced at each site is bounded by $J > 0$:
\begin{align}
    \sup_{x \in \Lambda} \left( \sum_{X \ni x} \norm{h_X}_{\infty} \right) \leq J. \label{eq:J_defn}
\end{align}
We define the boundary of a qudit region $A \subseteq \Lambda$ with respect to the interactions in $H$:
\begin{align}
 \partial A := \{ a \in A : \exists X \in \mathcal{E}, \; a \in X, \; X \cap (\Lambda \setminus A) \not= \varnothing \}.\label{eq:interaction_boundary_defn}
\end{align}
In words, $\partial A$ contains every site in $A$ touched by an interaction term $h_X$ which also has some support outside of $A$.

For a state $\rho$ defined on $\mathcal{H}$ and any subset of qudits $L \subseteq \Lambda$, the reduced state of $\rho$ on $L$ is defined as $\rho_L := \Tr_{\Lambda \setminus L}(\rho)$. 
In subscripts, we write $AC = A\sqcup C$, $ABC = A \sqcup B \sqcup C$, etc.
For two qudit subsets $A \sqcup C \subset \Lambda$, mutual information (MI) is evaluated in the reduced state $\rho_{AC}$:
\begin{align}
 I(A:C)_{\rho} = S(\rho_A)+S(\rho_C)-S(\rho_{AC}). \label{eq:MI_defn}
\end{align}
For a tripartition $A \sqcup B \sqcup C \subset \Lambda$, denote its complement by $D=\Lambda\setminus(A\sqcup B\sqcup C)$.
Conditional mutual information (CMI) is evaluated in $\rho_{ABC}=\Tr_D( \rho)$:
\begin{align}
 I(A:C | B)_{\rho} = S(\rho_{AB}) + S(\rho_{BC})-S(\rho_B)-S(\rho_{ABC}). \label{eq:CMI_defn}
\end{align}

\subsection{Main results}\label{subsec:main_results}
We now state our two main results on the decay of MI and CMI for high-temperature Gibbs states, followed by a summary of the approaches to their proofs.
\begin{theorem}[Exponential decay of mutual information]\label{thm:MI_decay}
In the setting of Sec.~\ref{subsec:setting}, let $\rho_\beta$ be the Gibbs state of a local Hamiltonian $H$ at inverse temperature $\beta$.
Define
\begin{align}
 \varepsilon_{\mathrm{MI}}=\frac{1}{8e^3k(2q-1)^k}. \label{eq:MI_decay_threshold_thm}
\end{align}
If $0\le\beta J<\varepsilon_{\mathrm{MI}}$, then for all $A \sqcup C \subseteq \Lambda$,
\begin{align}
 I(A:C)_{\rho_{\beta}} \leq K_{\mathrm{MI}} \, \min\{|\partial A|,|\partial C|\} \, e^{-\dist(A,C)/R} \label{eq:MI_decay_thm}
\end{align}
where $K_{\mathrm{MI}} = 8\beta J(2q-1)^k e^{\beta J+2}$.
\end{theorem}

\begin{theorem}[Exponential decay of conditional mutual information]\label{thm:CMI_decay}
In the setting of Sec.~\ref{subsec:setting}, let $\rho_\beta$ be the Gibbs state of a local Hamiltonian $H$ at inverse temperature $\beta$.
Define
\begin{align}
 \varepsilon_{\mathrm{CMI}} = \frac{1}{8ek(2q)^k e^{2k+1}}. \label{eq:CMI_decay_threshold_thm}
\end{align}
If $0\le\beta J\le\varepsilon_{\mathrm{CMI}}$, then for every $A \sqcup B \sqcup C \subseteq \Lambda$,
\begin{align}
 I(A:C | B)_{\rho_{\beta}} \leq K_{\mathrm{CMI}} \, \min\{|\partial A|,|\partial C|\} \, e^{-\dist(A,C)/R}, \label{eq:CMI_decay_thm}
\end{align}
where $K_{\mathrm{CMI}} = 2\log(q)  \mathsf{S}_{R/2}^{2} \mathsf{S}_R$, with $\mathsf{S}_R$ defined by Eq.~\eqref{eq:lattice_kernel}.
\end{theorem}

Theorems~\ref{thm:MI_decay} and~\ref{thm:CMI_decay} both rely on a classical cluster expansion of the Gibbs weights from our on-site 2-design decomposition.
Theorem~\ref{thm:CMI_decay} requires a stronger decay estimate on the resulting effective classical potentials. 
The proofs proceed in three main steps:
\begin{enumerate}
\item \textbf{Positive classical representation and cluster expansion (Section~\ref{sec:classical}).} 
We decompose the Gibbs state $\rho_\beta$ on each site using a local projective 2-design, expressing $\rho_\beta$ and all of its marginals $\rho_{\beta,L}$ (for $L \subseteq \Lambda$) as mixtures of tensor products of rank-1 projectors weighted by scalar functions $p_L(\s_L)$ where $\s_L$ labels the projectors. 
Performing a classical cluster expansion on these scalar weights shows that, at sufficiently high temperature, $p_L(\s_L)$ is strictly positive (recovering separability of $\rho_\beta$) and admits an absolutely convergent series for $\log(p_L(\s_L))$ uniform in the subsystem $L$.

\item \textbf{Decay of MI via cluster cancellation (Section~\ref{sec:MI_proof}).} 
Using the expansion for $\log(p_L(\s_L))$, we bound the classical log-likelihood ratio $\log \left[\frac{p_{AC}(\s_A,\s_C)}{p_A(\s_A)p_C(\s_C)}\right]$. 
All clusters that do not intersect both $A$ and $C$ cancel identically, leaving only connected clusters that cross both $\partial A$ and $\partial C$ and span $\dist(A,C)$. 
Because cluster weights decay exponentially with size, this yields an upper bound $\Delta_{AC}$ on the log-likelihood ratio which we can further lift directly to the operator inequality $\rho_{\beta,AC}\le e^{\Delta_{AC}}(\rho_{\beta,A}\otimes \rho_{\beta,C})$ and then obtain the exponential decay of MI bound via operator monotonicity of the logarithm.

\item \textbf{Decay of CMI via local resampling contraction (Section~\ref{sec:CMI_proof}).} 
To bound CMI, we regroup the cluster expansion for $\log(p(\s))$ by cluster support, representing the classical weights as a Gibbs distribution with potentials $\Psi_U(\s_U)$ that decay exponentially in both $|U|$ and $\diam(U)$. 
This decay allows us to decompose the conditional distribution of each label $s_x$ given its environment as a convex mixture of an independent sampling rule (with probability at least $1/2$) and local resampling rules.
Data processing under this resampling yields a contraction recurrence for the CMI between two sites while keeping the conditioning registers quantum, implying exponential decay of pairwise CMI in $\dist(a,c)$. 
Finally, applying this estimate on the interaction graph and factoring shortest paths through $\partial A$ and $\partial C$ converts the sum over site pairs into the boundary prefactor $\min\{|\partial A|,|\partial C|\}$.
\end{enumerate}

\section{Classical representation of quantum Gibbs states}\label{sec:classical}
\subsection{On-site projective 2-design decomposition}
To begin, we define the projective 2-designs used to decompose the Gibbs state. 
For any $x \in \Lambda$, let $\mathcal{S}_x$ be a finite set of classical labels, and let $\nu_x$ be a positive probability distribution on the labels:
\begin{align}
    \nu_x(s_x) > 0 \text{ for each } s_x \in \mathcal{S}_x, \text{ with } \sum_{s_x \in \mathcal{S}_x}\nu_x(s_x) = 1.
\end{align}
Each label $s_x$ is associated with a distinct rank-1 projector $P_{s_x}$ on $\mathcal{H}_x$. 
The set of projectors need not be orthogonal---in fact, they will form an overcomplete basis for operators acting on $\mathcal{H}_x$.
This is because we require that the ensemble $\{(\nu_x(s_x),P_{s_x}) : s_x \in \mathcal{S}_x\}$ formed by the labels, probabilities, and projectors is a 2-design: its first two moments must agree with those of a Haar-random pure state \cite{Mele2024Haar}. 
Writing $\mathsf{Swap}$ for the operator that exchanges states on two copies of $\mathcal H_x$, this amounts to imposing that
\begin{align}
 \underset{s_x \sim \nu_x}{\E}\left[ P_{s_x} \right]=\frac{\id}{d_x},
 \qquad 
 \underset{s_x \sim \nu_x}{\E}\left[P_{s_x,1}\otimes P_{s_x,2} \right] =\frac{\id_1 \otimes \id_2+\mathsf{Swap}_{12}}{d_x(d_x+1)}. \label{eq:2-design_moments}
\end{align}
Such a projective 2-design always exists when $d_x$ is finite; in particular, it can be formed from a finite union of orthonormal bases \cite[Corollary~3.2]{RoyScott2007Designs}.

For each label $s_x$, define the dual of its projector $P_{s_x}$ as
\begin{align}
    D_{s_x} := (d_x+1)P_{s_x}-\id.
\end{align}
Then, for any operator $O_x$ acting on $\mathcal H_x$, define
\begin{align}
 \tau_{s_x}(O_x) := \Tr_x[D_{s_x}O_x], \qquad \tr_x(O_x) := \frac{\Tr_x[O_x]}{d_x}, \label{eq:local_functionals}
\end{align}
where $\tau_{s_x}(O_x)$ is the `weight of $O_x$ along $P_{s_x}$' and $\tr_x(O_x)$ is the normalized trace over $\mathcal{H}_x$.
For a linear functional $\tau$ on operators, denote the dual norm $\norm{\tau}=\sup_{\norm{O}_\infty \leq 1}|\tau(O)|$.
The projectors being rank-1 and the moment identities in Eq. \eqref{eq:2-design_moments} give
\begin{align}
 \tau_{s_x}(\id)=1,\qquad
 \norm{\tau_{s_x}}=2d_x-1\le 2q-1,\qquad
 \underset{s_x \sim \nu_x}{\E}[\tau_{s_x}(\cdot)]=\tr_x(\cdot). \label{eq:local_functional_properties}
\end{align}
We can then decompose any local operator $O_x$ as a weighted sum of the rank-1 projectors $\left\{ P_{s_x} \right\}$:
\begin{align}
    O_x = d_x \underset{s_x \sim \nu_x}{\E}\big[\tau_{s_x}(O_x) P_{s_x}\big].\label{eq:local_operator_decomposition}
\end{align}
To see this, take the partial trace over the second copy in the second moment identity in Eq.~\eqref{eq:2-design_moments} after multiplying by $\id \otimes O_x$. 
This gives
\begin{align}
 \underset{s_x \sim \nu_x}{\E}\left[P_{s_x}\Tr_x(P_{s_x}O_x)\right]
 =\frac{O_x+\Tr_x(O_x)\id}{d_x(d_x+1)}.
\end{align}
Using the definition of $\tau_{s_x}$ and the first moment identity, we therefore obtain
\begin{align}
 d_x\E_{\nu_x}\!\left[\tau_{s_x}(O_x)P_{s_x}\right]
 &=d_x(d_x+1)\E_{\nu_x}\!\left[P_{s_x}\Tr_x(P_{s_x}O_x)\right]
   -d_x \Tr_x(O_x) \cdot \E_{\nu_x}[P_{s_x}] \nonumber \\
 &=O_x +\Tr_x(O_x)\id - \Tr_x (O_x)\id = O_x.
\end{align}

Since the local Hilbert space at each $x \in \Lambda$ is finite-dimensional with $d_x \leq q$, we can use a projective 2-design decomposition at each site to represent arbitrary operators $O$ acting on $\mathcal{H}$.
Denote the set of global labels by $\mathcal{S} := \prod_{x \in \Lambda} \mathcal{S}_x$, writing its elements as $\s = (s_x)_{x\in\Lambda} \in \mathcal{S}$.
The full (product) probability distribution on the global labels in $\mathcal{S}$ is $\nu(\s) := \prod_{x \in \Lambda}\nu_x(s_x)$.
Now, we can decompose any operator $O$ acting on $\mathcal{H}$ as
\begin{align}
    O = \left( \prod_{x \in \Lambda} d_x \right) \underset{\s \sim \nu}{\E}\Bigg[ \omega_{\s}(O) \bigotimes_{x \in \Lambda} P_{s_x} \Bigg], \label{eq:global_operator_decomposition}
\end{align}
where we have defined a linear product functional on global operators as
\begin{align}
    \omega_\s(O) := \left(\bigotimes_{x \in \Lambda} \tau_{s_x}\right) (O)= \Tr\left\{ \left( \bigotimes_{x \in \Lambda} D_{s_x} \right) O \right\}. \label{eq:product_functional}
\end{align}
By Eq.~\eqref{eq:local_functional_properties}, the functional $\omega_{\s}$ is unital and has norm at most $(2q-1)^{|U|}$ on operators supported in $U$.
Furthermore, it is multiplicative over disjoint supports:
\begin{align}
    \omega_{\s}(O_A \otimes O_B) = \omega_{\s}(O_A \otimes \id_B) \cdot \omega_{\s}(\id_A \otimes O_B) \;\text{ for }\; A \sqcup B = \Lambda. \label{eq:functional_factorization}
\end{align}

Applying the decomposition in Eq.~\eqref{eq:global_operator_decomposition} to the Gibbs operator $e^{-\beta H}$ gives
\begin{align}
    e^{-\beta H} = \left( \prod_{x \in \Lambda} d_x \right) \underset{\s \sim \nu}{\E}\Bigg[ \omega_{\s}(e^{-\beta H}) \bigotimes_{x \in \Lambda} P_{s_x} \Bigg], \label{eq:Gibbs_operator_2-design_decomposition}
\end{align}
where we will refer to each coefficient $\omega_\s(e^{-\beta H})$ as a \textit{Gibbs weight} (of a given label $\s$).
The notation and decomposition introduced for global operators and the full Gibbs state carry over to operators supported on a subset of qudits $L \subseteq \Lambda$. 
The set of global labels restricted to $L$ is written as $\mathcal{S}_L := \prod_{x \in L} \mathcal{S}_x$, its elements as $\s_L=(s_x)_{x\in L}$, and its distribution as $\nu_L(\s_L)=\prod_{x\in L}\nu_x(s_x)$.
For any fixed $\s_L \in \mathcal{S}_L$, define the weight functional as
\begin{align}
 \omega_{\s_L} := \bigotimes_{x\in L}\tau_{s_x} \otimes\bigotimes_{x \in L^c} \tr_x \label{eq:restricted_product_functional}
\end{align}
where $L^c := \Lambda \setminus L$ is the complement of $L$.
Although $\omega_{\s_L}$ is defined to act on global operators, the restriction of its label $\s_L$ indicates that it represents `measurements' on $L$ alone. 
Applying this functional to the Gibbs operator gives the marginal Gibbs weight $\omega_{\s_L}(e^{-\beta H})$.
For $L=\varnothing$, we write $\omega_\varnothing=\bigotimes_{x\in\Lambda}\tr_x$. 
By Eq.~\eqref{eq:local_functional_properties}, averaging over the local label at a site $x$ replaces $\tau_{s_x}(\cdot)$ by $\tr_x(\cdot)$, so
\begin{align}
 \underset{\s_{L^c} \sim \nu_{L^c}}{\E}\left[\omega_{\s}(e^{-\beta H})\right] &=\omega_{\s_L}(e^{-\beta H}), \qquad
 \omega_\varnothing(e^{-\beta H}) =\frac{\Tr(e^{-\beta H})}{\prod_xd_x} = \frac{Z_\beta}{\prod_x d_x}, \label{eq:marginal_Gibbs_weight}
\end{align}
and the reduced Gibbs operator on a qudit region $L \subseteq \Lambda$ is
\begin{align}
    \left( e^{-\beta H} \right)_L = \Tr_{L^c}\left( e^{-\beta H} \right) = \left( \prod_{x \in \Lambda} d_x \right) \underset{\s_L \sim \nu_L}{\E}\Bigg[ \omega_{\s_L}(e^{-\beta H}) \bigotimes_{x \in L} P_{s_x} \Bigg]. \label{eq:Gibbs_marginal_decomposition}
\end{align}
In this sense, the 2-design decomposition `commutes with the partial trace', allowing us to represent the full and reduced Gibbs state using a common expansion.
To express the full Gibbs state, we note that the Gibbs weight $\omega_{\varnothing}(e^{-\beta H})$ in Eq.~\eqref{eq:marginal_Gibbs_weight} gives the partition function, so we can normalize the decomposed Gibbs operator in Eq.~\eqref{eq:Gibbs_operator_2-design_decomposition} as
\begin{align}
    \rho_\beta = \underset{\s \sim \nu}{\E} \left[ \frac{\omega_\s(e^{-\beta H})}{\omega_{\varnothing}(e^{-\beta H})} \bigotimes_{x \in \Lambda} P_{s_x} \right]. \label{eq:Gibbs_state_2-design_decomposition}
\end{align}
We note that, as the decomposition in Eq.~\eqref{eq:global_operator_decomposition} applies to arbitrary operators, $\omega_\s$ is not a positive functional, so the scalar Gibbs weights $\omega_\s(e^{-\beta H})$ are not strictly positive in general.
However, looking at Eq.~\eqref{eq:Gibbs_state_2-design_decomposition}, one can see that proving each Gibbs weight is strictly positive $\omega_\s(e^{-\beta H}) > 0$ immediately implies the separability of the Gibbs state $\rho_\beta$.

\subsection{Cluster expansion}
We now perform a standard cluster expansion of the quantum Gibbs operator $e^{-\beta H}$.
Define a \textit{polymer} $\gamma = \left\{X_1,\ldots,X_{|\gamma|} \right\}$ to be a nonempty, unordered set of interaction supports $X_i \in \mathcal E$ whose overlap graph is connected: the graph has one vertex for each $X_i$ and an edge between distinct vertices $i,j$ whenever $X_i \cap X_j \not= \varnothing$. 
Denote the total support of a polymer $\gamma$ as
\begin{align}
 U_\gamma=\bigcup_{X\in\gamma} X. \label{eq:polymer_support}
\end{align}
Two polymers $\gamma,\gamma'$ are \textit{incompatible} when their supports intersect, i.e. when $U_{\gamma} \cap U_{\gamma'} \not= \varnothing$, so every polymer is incompatible with itself.
If two polymers are not incompatible, then they are \textit{compatible}.
A \textit{word} of length $n$ is an ordered tuple $(X_1,\ldots,X_n)\in\mathcal E^n$ which is to be associated with the operator product $h_{X_1} \cdots h_{X_n}$.
Repetitions of interaction supports are allowed within a word.
Package all repetitions of the labels in
$\gamma$ into the operator
\begin{align}
 Q_\gamma:=
 \sum_{n\ge |\gamma|}\frac{(-\beta)^n}{n!}
 \sum_{\substack{(X_1,\ldots,X_n)\in\gamma^n\\
                 \{X_1,\ldots,X_n\}=\gamma}}
 h_{X_1}\cdots h_{X_n}. \label{eq:Q_definition}
\end{align}
The inner sum runs over ordered tuples of interaction labels in
$\gamma$, with every label appearing at least once. 
Let $\mathcal G$ range over unordered collections of pairwise compatible polymers, meaning that their site supports are pairwise disjoint.
In the Taylor expansion of $e^{-\beta H}$, grouping the distinct interaction labels in each word into connected components under support overlap and summing over all ways to interleave them gives
\begin{align}
 e^{-\beta H} =\sum_{\substack{\mathcal{G} \; \mathrm{pairwise} \\ \mathrm{compatible}}} \prod_{\gamma\in\mathcal G}Q_\gamma. \label{eq:operator_gas}
\end{align}
Grouping words in $Q_\gamma$ by their multiplicities gives
\begin{align}\label{eq:polymer-operator-bound}
 \norm{Q_\gamma}_\infty
 \le\prod_{X\in\gamma}(e^{\beta\norm{h_X}_\infty}-1).
\end{align}
For any $L \subseteq \Lambda$ and label $\s_L \in \mathcal{S}_L$, applying $\omega_{\s_L}$ to $Q_\gamma$ gives a scalar-valued
function of the labels, with the uniform bound
\begin{align}\label{eq:polymer-scalar-bound}
 |\omega_{\s_L}(Q_\gamma)|
 \le\prod_{X\in\gamma}(2q-1)^k(e^{\beta\norm{h_X}_\infty}-1).
\end{align}
Applying $\omega_{\s_L}$ to the expression in Eq.~\eqref{eq:operator_gas}, using the factorization over disjoint supports in Eq.~\eqref{eq:functional_factorization}, and using the linearity of $\omega_{\s_L}$ gives
\begin{align}
    \omega_{\s_L}(e^{-\beta H}) = \sum_{\substack{\mathcal{G} \; \mathrm{pairwise} \\ \mathrm{compatible}}} \prod_{\gamma\in\mathcal G}\omega_{\s_L}(Q_\gamma). \label{eq:polymer_gas}
\end{align}

For a \textit{cluster} (i.e. an ordered tuple of polymers) $\Gamma=(\gamma_1,\ldots,\gamma_n)$ with $n\ge1$, in which the same polymer may occur more than once, write
\begin{align}\label{eq:cluster-support}
 U_\Gamma=\bigcup_{j=1}^nU_{\gamma_j},\qquad
 N_\Gamma=\sum_{j=1}^n|\gamma_j|.
\end{align}
Let $[n]=\{1,\ldots,n\}$. For a simple graph $G$ on $[n]$, write
$E(G)$ for its edge set, and let $\mathbf{1}_{\{\cdot\}}$ denote an indicator.
Define
\begin{align}\label{eq:truncated-coefficient}
 \phi^{\mathsf T}(\gamma_1,\ldots,\gamma_n)
 =
 \sum_{\substack{G\ {\rm connected}\\\text{on }[n]}}
 \prod_{\{i,j\}\in E(G)}
       \big(-\mathbf1_{U_{\gamma_i}\cap U_{\gamma_j}\ne\varnothing}\big),
\end{align}
with $\phi^{\mathsf T}(\gamma_1)=1$.
The incompatibility graph of $\Gamma$ joins distinct vertices $i,j$
when $U_{\gamma_i}\cap U_{\gamma_j}\ne\varnothing$; the coefficient
vanishes unless this graph is connected.

\begin{lemma}[Linked-cluster theorem {\cite[Section~2]{Ueltschi}}]\label{lem:linked-cluster}
Fix a qudit region $L \subseteq \Lambda$ and label $\s_L \in \mathcal{S}_L$. 
If the connected series on the right-hand side of Eq.~\eqref{eq:linked_cluster} converges absolutely, then the Gibbs weight $\omega_{\s_L}(e^{-\beta H})$ is positive and its logarithm satisfies
\begin{align}
 \log \Bigg(
 \sum_{\substack{\mathcal{G} \; \mathrm{pairwise} \\ \mathrm{compatible}}} \prod_{\gamma\in\mathcal G} \omega_{\s_L}(Q_\gamma)\Bigg) = 
 \sum_{n\ge1}\frac1{n!} \sum_{\gamma_1,\ldots,\gamma_n} \phi^{\mathsf{T}}(\gamma_1,\ldots,\gamma_n) \prod_{j=1}^n\omega_{\s_L}(Q_{\gamma_j}). \label{eq:linked_cluster}
\end{align}
\end{lemma}

\begin{lemma}[Koteck\'y--Preiss {\cite{KP,Ueltschi}}]\label{lem:kp}
Suppose nonnegative $z_\gamma,a_\gamma$ satisfy,
for every polymer $\gamma$,
\begin{align}\label{eq:kp_condition}
 \sum_{\gamma':\,U_{\gamma'}\cap U_\gamma\ne\varnothing}
            z_{\gamma'}e^{a_{\gamma'}}\le a_\gamma.
\end{align}
Then, for any $L \subseteq \Lambda$ and $\s_L \in \mathcal{S}_L$ such that $|\omega_{\s_L}(Q_\gamma)|\le z_\gamma$ for every polymer $\gamma$, the connected series in Lemma~\ref{lem:linked-cluster} converges absolutely and exponentiates to $\omega_{\s_L}(e^{-\beta H})$.
Moreover,
\begin{align}\label{eq:kp_rooted_bound}
 \sum_{n\ge1}\frac1{(n-1)!}\sum_{\gamma_2,\ldots,\gamma_n}
 |\phi^{\mathsf T}(\gamma,\gamma_2,\ldots,\gamma_n)|
 \prod_{j=2}^n z_{\gamma_j}\le e^{a_\gamma}.
\end{align}
For every site $x \in \Lambda$, the following bound also holds:
\begin{align}\label{eq:kp_site}
 \sum_{n\ge1}\frac1{n!}
 \sum_{\substack{\gamma_1,\ldots,\gamma_n\\
                 x\in\cup_jU_{\gamma_j}}}
 |\phi^{\mathsf T}(\gamma_1,\ldots,\gamma_n)|\prod_jz_{\gamma_j}
 \le\sum_{\gamma:\,x\in U_\gamma}z_\gamma e^{a_\gamma}.
\end{align}
\end{lemma}

The connected expansion in Lemma~\ref{lem:linked-cluster} reads
\begin{align}
 \log \left[ \omega_{\s_L}(e^{-\beta H}) \right]
 =\sum_{n\ge1}\frac1{n!}\sum_{\gamma_1,\ldots,\gamma_n}
       \phi^{\mathsf{T}}(\gamma_1,\ldots,\gamma_n)
       \prod_j\omega_{\s_L}(Q_{\gamma_j}). \label{eq:log_weight_expansion}
\end{align}
For each nonempty $U\subseteq\Lambda$, define a `classical potential' $\Psi_U$ as the series
\begin{align}\label{eq:potential-definition}
 \Psi_U(\s_U)=\sum_{n\ge1}\frac1{n!}
 \sum_{\substack{\gamma_1,\ldots,\gamma_n\\U_\Gamma=U}}
 \phi^{\mathsf T}(\gamma_1,\ldots,\gamma_n)
 \prod_{j=1}^n\omega_{\s}(Q_{\gamma_j}).
\end{align}
Here $U_\Gamma=U$ restricts the sum to tuples whose interaction supports' union is exactly $U$. 
Each term depends only on $\s_U$. 
Grouping by this union gives
\begin{align}\label{eq:classical-potential}
 \log \left[\omega_{\s}(e^{-\beta H}) \right] = \sum_{U \not= \varnothing} \Psi_U(\s_U)
\end{align}
The following lemma establishes absolute convergence and justifies
these identities.

\begin{lemma}[Convergent cluster expansion for Gibbs weights]\label{lem:expansion}
For $\lambda\ge0$, put
\begin{align}
 \eta_\lambda=\beta J(2q-1)^k e^{\beta J+\lambda+1}. \label{eq:eta_lambda_defn}
\end{align}
If $\eta_\lambda\le1/(8k)$, then
$\omega_{\s_L}(e^{-\beta H})>0$ for every $L \subseteq \Lambda$ and $\s_L \in \mathcal{S}_L$.
The series in \eqref{eq:log_weight_expansion} and \eqref{eq:potential-definition}
converge absolutely and uniformly in their labels, with uniformity in
$L$ for \eqref{eq:log_weight_expansion}. Equations~\eqref{eq:log_weight_expansion} and
\eqref{eq:classical-potential} hold with the real logarithm,
and the functions $\Psi_U$ are real-valued.
Moreover,
\begin{align}\label{eq:root_bound}
 \sup_x\sum_{\Gamma:\,x\in U_\Gamma}
             \frac{|\phi^{\mathsf T}(\Gamma)|}{n!}
             \left(\prod_{j=1}^n\prod_{X\in\gamma_j}(2q-1)^k(e^{\beta\norm{h_X}_\infty}-1)\right)
             e^{\lambda N_\Gamma}\le2\eta_\lambda.
\end{align}
In each summand, $\Gamma=(\gamma_1,\ldots,\gamma_n)$; the tuple sum
includes all lengths $n\ge1$ and repetitions with the displayed factorials.

If $\eta_{k+2}\le1/(8k)$, let $\delta$ be a metric on $\Lambda$ and
let $r>0$ satisfy $\diam_\delta(X)\le r$ for every $X\in\mathcal E$.
Then, with $\|\Psi_U\|_\infty=\max_{s_U}|\Psi_U(s_U)|$,
\begin{align}\label{eq:potential}
 \sup_x\sum_{U\ni x}
    e^{|U|+2\diam_\delta(U)/r}\norm{\Psi_U}_\infty
 \leq 2\eta_{k+2}.
\end{align}
\end{lemma}

\begin{proof}
The parameter $\lambda$ reserves an exponential weight for the final estimate. 
Set
\begin{align}\label{eq:weighted-interaction}
 v_X=e^{\lambda+1}(2q-1)^k(e^{\beta\norm{h_X}_\infty}-1)
\end{align}
for any $X \in \mathcal{E}$.
Using
$\norm{h_X}_\infty\le J$ and $e^x-1\le xe^x$ for $x\ge0$ gives
\begin{align}\label{eq:weighted_interaction_sum}
 \sup_x\sum_{X\ni x}v_X
 \le e^{\lambda+1}(2q-1)^k\beta e^{\beta J}
             \sup_x\sum_{X\ni x}\norm{h_X}_\infty
 \le\eta_\lambda.
\end{align}
For each $m\ge1$, counting polymers with $m$ interaction labels by rooted trees gives
\begin{align}\label{eq:rooted_tree_count}
 \sum_{\substack{\gamma:\,x\in U_\gamma\\|\gamma|=m}}
       \prod_{X\in\gamma}v_X
 \le\eta_\lambda(4k\eta_\lambda)^{m-1}.
\end{align}
Indeed, choose a rooted ordered spanning tree for each connected set, rooted at a label containing $x$.
There are at most $4^{m-1}$ plane rooted tree shapes.
The root-label sum costs at most $\eta_\lambda$; each child-label sum costs at most $k\eta_\lambda$, because it must overlap its parent. 
Allowing repeated labels in this upper bound only increases the sum. 
Appendix~\ref{app:rooted_tree_count} gives the detailed counting and summation argument.

It follows that
\begin{align}\label{eq:polymer-site-sum}
 \sum_{\gamma:\,x\in U_\gamma}
                  e^{(\lambda+1)|\gamma|}\prod_{X\in\gamma}(2q-1)^k(e^{\beta\norm{h_X}_\infty}-1)
 \leq \frac{\eta_\lambda}{1-4k\eta_\lambda}
 \leq 2\eta_\lambda
\end{align}
where we have used $\eta_\lambda \leq 1/(8k)$ for the last inequality. 
Applying Lemma~\ref{lem:kp} with
\begin{align}
 z_\gamma=e^{\lambda |\gamma|}\prod_{X\in\gamma}(2q-1)^k(e^{\beta\norm{h_X}_\infty}-1), \qquad a_\gamma=|\gamma|, \label{eq:kp-parameters}
\end{align}
we have that
\begin{align}
 \sum_{\gamma':\,U_{\gamma'} \cap U_\gamma \not= \varnothing} z_{\gamma'}e^{a_{\gamma'}}
 \leq \sum_{x\in U_\gamma} \sum_{\gamma':\,x\in U_{\gamma'}}z_{\gamma'}e^{a_{\gamma'}}
 \leq 2k\eta_\lambda |\gamma|\le |\gamma|. \label{eq:kp-verification}
\end{align}
Lemma~\ref{lem:kp} then gives absolute convergence of the connected series expansion of $\log[\omega_{\s_L}(e^{-\beta H})]$ from Lemma~\ref{lem:linked-cluster}.
Eq.~\eqref{eq:kp_site} of Lemma~\ref{lem:kp} proves \eqref{eq:root_bound}, and Lemma~\ref{lem:linked-cluster} identifies its exponential with $\omega_{\s_L}(e^{-\beta H})$.
Hence $\omega_{\s_L}(e^{-\beta H}) > 0$.

For the potential bound, group the absolutely convergent series in Eq.~\eqref{eq:log_weight_expansion} by its exact union $U_\Gamma$, as in Eq.~\eqref{eq:potential-definition}.
A nonzero cluster is connected through overlapping interaction supports. 
Hence
\begin{align}\label{eq:cluster-size-diameter}
 |U_\Gamma|\le kN_\Gamma,\qquad
 \diam_\delta(U_\Gamma)\le rN_\Gamma.
\end{align}
The triangle inequality bounds $\|\Psi_U\|_\infty$ by the sum of the absolute cluster bounds with union $U$.
The desired weight for each cluster is at most $e^{(k+2)N_\Gamma}$, so \eqref{eq:root_bound} with $\lambda=k+2$ proves Eq.~\eqref{eq:potential}.
\end{proof}

Thus, the cluster expansion uniformly converges for all Gibbs weights $\omega_\s(e^{-\beta H})$ (and marginal Gibbs weights) at sufficiently small $\beta$, so the on-site 2-design decomposition yields a positive representation of $\rho_\beta$:

\begin{corollary}[Positive local preparation]\label{cor:preparation}
Under the hypothesis of Lemma~\ref{lem:expansion},
\begin{align}
 p(\s)=\frac{\omega_{\s}(e^{-\beta H})}{\omega_\varnothing(e^{-\beta H})} \label{eq:classical-density}
\end{align}
is a strictly positive probability density relative to $\nu$, defining the classical probability law $\pi(\s)=p(\s)\nu(\s)$. 
Its marginal density on $L \subseteq \Lambda$ is
\begin{align}
 p_L(\s_L)=\frac{\omega_{\s_L}(e^{-\beta H})}
 {\omega_\varnothing(e^{-\beta H})}. \label{eq:classical_marginal_density}
\end{align}
The corresponding quantum states satisfy
\begin{align}
 \rho_\beta = \underset{\s \sim \nu}{\E} \left[ p(\s)\bigotimes_{x \in \Lambda} P_{s_x}\right],
 \qquad
 \rho_{\beta,L}=\underset{\s_L \sim \nu_L}{\E} \left[ p_L(\s_L) \bigotimes_{x\in L}P_{s_x} \right]. \label{eq:positive_preparation}
\end{align}
\end{corollary}

Since the Gibbs state $\rho_\beta$ in Eq.~\eqref{eq:positive_preparation} is a convex mixture of product states, it is fully separable below a constant value of $\beta$.
This provides an alternate derivation of the sudden death of thermal entanglement~\cite{BakshiLiuMoitraTang2024} below a constant inverse temperature threshold.

\section{Mutual information: cancellation leaves spanning clusters}
\label{sec:MI_proof}

The key quantity for proving exponential decay of the MI is a scalar likelihood ratio.
Its logarithm is a difference of four marginal logarithms, in which clusters supported away from either endpoint cancel exactly.

\begin{proof}[Proof of Theorem~\ref{thm:MI_decay}]
We assume the temperature threshold in Eq.~\eqref{eq:MI_decay_threshold_thm}, which implies that $\beta J<1$ and that, setting $\lambda = 1$ for $\eta_\lambda$ defined in Eq.~\eqref{eq:eta_lambda_defn}, 
\begin{align}\label{eq:mi-temperature}
 \eta_1=\beta J(2q-1)^k e^{\beta J+2}<\frac1{8k}.
\end{align}
Thus, the uniform convergence of the Gibbs weight cluster expansions in Lemma~\ref{lem:expansion} and the positive preparation of Corollary~\ref{cor:preparation} hold with $8\eta_1=K_{\mathrm{MI}}$.
Pointwise in the label set $\mathcal{S}_{AC}$, define
\begin{align}
 \ell(\s_A,\s_C)
 &=\log \left[\frac{p_{AC}(\s_A,\s_C)}{p_A(\s_A)p_C(\s_C)}\right] \\
 &=\log \left[\omega_{(\s_A,\s_C)}(e^{-\beta H})\right]
   +\log \left[\omega_\varnothing(e^{-\beta H})\right] -\log \left[\omega_{\s_A}(e^{-\beta H})\right]
   -\log \left[\omega_{\s_C}(e^{-\beta H})\right]. \label{eq:likelihood}
\end{align}
For a fixed cluster $\Gamma$, its contribution depends only on the local functionals at sites in $U_\Gamma$.
If $U_\Gamma$ misses $C$, the $AC$ and $A$ contributions agree, as do the $C$ and $\varnothing$ contributions.
The same cancellation holds if it misses $A$.
Absolute convergence therefore leaves only clusters meeting both endpoint regions.

A surviving connected cluster contains an interaction path from $A$ to $C$.
That path crosses both interaction boundaries, and its length gives
\begin{align}\label{eq:spanning-cluster}
 U_\Gamma\cap\partial A\ne\varnothing,\qquad
 U_\Gamma\cap\partial C\ne\varnothing,\qquad
 RN_\Gamma\ge \dist(A,C).
\end{align}
Each of the four contributions in Eq.~\eqref{eq:likelihood} is bounded by
\begin{align}\label{eq:cluster-contribution}
 \frac{|\phi^{\mathsf T}(\Gamma)|}{n!}
 \prod_{j=1}^n\prod_{X\in\gamma_j}(2q-1)^k(e^{\beta\norm{h_X}_\infty}-1).
\end{align}
By symmetry, assume $|\partial A|\le|\partial C|$.
For each surviving cluster, Eqs.~\eqref{eq:spanning-cluster} gives $e^{-N_\Gamma}\le e^{-\dist(A,C)/R}$.
Inserting $1=e^{-N_\Gamma}e^{N_\Gamma}$ into its weight therefore extracts the exponential decay factor.
Since every such cluster meets $\partial A$, we can bound the remaining sum by summing over a boundary site $a\in\partial A$ and all clusters containing $a$.
Finally, Eq.~\eqref{eq:root_bound} with $\lambda=1$ bounds each site-anchored sum by $2\eta_1$, and $8\eta_1$ is exactly $K_{\mathrm{MI}}$. 
Thus,
\begin{align}\label{eq:mi-cluster-bound}
 |\ell|
 &\le4\sum_{\substack{\Gamma:\,U_\Gamma\cap A\not=\varnothing, \\
                       U_\Gamma\cap C\not=\varnothing}}
       \frac{|\phi^{\mathsf T}(\Gamma)|}{n!}\prod_{j=1}^n\prod_{X\in\gamma_j}(2q-1)^k(e^{\beta\norm{h_X}_\infty}-1)
\\
 &\le4e^{-\dist(A,C)/R}
       \sum_{a\in\partial A}
       \sum_{\Gamma:\,a\in U_\Gamma}
       \frac{|\phi^{\mathsf T}(\Gamma)|}{n!}
       \left(\prod_{j=1}^n\prod_{X\in\gamma_j}(2q-1)^k(e^{\beta\norm{h_X}_\infty}-1)\right)e^{N_\Gamma}
\\
 &\le8\eta_1|\partial A|e^{-\dist(A,C)/R}=K_{\mathrm{MI}}|\partial A|e^{-\dist(A,C)/R}=: \Delta_{AC}.
\end{align}

To transfer this bound to a bound on the quantum MI, exponentiate the likelihood bound and use the positive preparations in Eq.~\eqref{eq:positive_preparation}:
\begin{align}\label{eq:likelihood-operator-bound}
 p_{AC}(\s_A,\s_C)\le e^{\Delta_{AC}}p_A(\s_A)p_C(\s_C)
 \quad\Longrightarrow\quad
 \rho_{\beta,AC}\le e^{\Delta_{AC}}(\rho_{\beta,A}\otimes\rho_{\beta,C}).
\end{align}
All reduced Gibbs states are positive definite.
Operator monotonicity of the logarithm therefore gives
\begin{align}\label{eq:logarithmic-operator-bound}
 \log\left[\rho_{\beta,AC}\right]-\log\left[\rho_{\beta,A}\otimes\rho_{\beta,C}\right]
 \le\Delta_{AC}\id.
\end{align}
Multiplying both sides by $\rho_{\beta,AC}$ and taking the trace proves Eq.~\eqref{eq:MI_decay_thm} and thus completes the proof.
\end{proof}

\section{Conditional mutual information: local contraction}
\label{sec:CMI_proof}
By Corollary~\ref{cor:preparation}, we can write the high-temperature Gibbs distribution on the global label set $\mathcal{S}$ as
\begin{align}
    \pi(\s) &= \nu(\s) \cdot \frac{\omega_\s(e^{-\beta H})}{\omega_{\varnothing}(e^{-\beta H})}  \propto \nu(\s) \exp \Bigg( \sum_{\varnothing \not= U \subseteq \Lambda} \Psi_U(\s) \Bigg). \label{eq:pi_full_distribution}
\end{align}
In this section, we will sometimes drop the explicit brackets for a set containing a single site, simply writing $x$ instead of $\{x\}$.
For any $x \in \Lambda$, denote $x^c := \Lambda \setminus \left\{ x \right\}$.
We can marginalize Eq.~\eqref{eq:pi_full_distribution} by conditioning on the label $\s_{x^c}$, giving
\begin{align}
    \pi(s_x | \s_{x^c}) = \frac{\pi(\s)}{\pi(\s_{x^c})} &= \nu_x(s_x) \cdot \frac{\exp\lrp{\sum_{\varnothing \not= U \subseteq \Lambda} \Psi_U(\s)}}{\underset{s_x \sim \nu_x}{\E}\lrb{\exp\lrp{\sum_{\varnothing \not= U \subseteq \Lambda} \Psi_U(\s)}}} \\
    &= \nu_x(s_x) \cdot \frac{\exp\lrp{\sum_{U \ni x} \Psi_U(\s)}}{\underset{s_x \sim \nu_x}{\E}\lrb{\exp\lrp{\sum_{U \ni x} \Psi_U(\s)}}} \label{eq:conditional_distribution_exp_ratio}
\end{align}
where equality in the second line is due to linearity of expectation. 
When the family of classical potentials defining this conditional distribution satisfies a weighted upper bound (which we prove to hold at sufficiently high temperature), we can rewrite the conditional probability for resampling a label $s_x$ at site $x$ given all other labels $\s_{x^c}$:
\begin{lemma}[Stochastic kernel decomposition]\label{lem:stochastic_kernel_decomposition}
Suppose that, for some $\mu > 0$, the family of classical potentials in Eq.~\eqref{eq:pi_full_distribution} satisfies
\begin{align}
    \mathfrak{g} := \underset{x \in \Lambda}{\sup} \sum_{U \ni x} |U| e^{\mu \diam(U)} \norm{\Psi_U}_\infty \leq \log(6/5). \label{eq:stochastic_kernel_assumption}
\end{align}
Then, for any $x \in \Lambda$, the conditional distribution in Eq.~\eqref{eq:conditional_distribution_exp_ratio} admits a stochastic kernel decomposition
\begin{align}
    \pi(s_x | \s_{x^c}) = \lambda_{x,0} \nu_x(s_x) + \sum_{U \ni x} \lambda_{x,U} \K_{x,U}(s_x | \s_{U \setminus x}) \label{eq:stochastic_kernel_decomposition}
\end{align}
where the weights $\lambda_{x,U}$ are label-independent, nonnegative, sum to $1$, and decay with $|U|$:
\begin{align}
    \sum_{U \ni x} |U| e^{\mu \diam(U)} \lambda_{x,U} \leq \frac{1}{2}.
\end{align}
Each kernel $\K_{x,U}(s_x | \s_{U \setminus x})$ is positive and normalized, and the weight of the independent resampling branch satisfies $\lambda_{x,0} \geq 1/2$, meaning that sampling $s_x$ independently of all other sites occurs with probability at least $1/2$.
\end{lemma}
Once a resampling law in the form of Eq.~\eqref{eq:stochastic_kernel_decomposition} is established, we will show that its corresponding pairwise influence matrix $\C_{xy}$ contracts the maximum pairwise CMI between any two sites.
This is used to establish the exponential decay of pairwise CMI between any two sites $a \in A$ and $c \in C$ conditioned on $B$.
Applying this, we obtain the following:
\begin{lemma}[Contraction of conditional mutual information]\label{lem:conditional_information_contraction}
Suppose that, for some $\mu > 0$, the family of classical potentials in Eq.~\eqref{eq:pi_full_distribution} satisfies the assumption of Lemma~\ref{lem:stochastic_kernel_decomposition} in Eq.~\eqref{eq:stochastic_kernel_assumption}.
Then, for any $A \sqcup B \sqcup C \subseteq \Lambda$, we have
\begin{align}
    I(A:C|B)_{\rho_\beta} \leq 2 \log(q) \sum_{a \in A,\, c \in C} e^{-\mu \dist(a,c)}.
\end{align}
\end{lemma}
Finally, we will convert the bound in Lemma~\ref{lem:conditional_information_contraction} to the bound stated in Theorem~\ref{thm:CMI_decay}.

\subsection{Constructing a local resampling law}
To prove Lemma~\ref{lem:stochastic_kernel_decomposition}, we first introduce some formalism to perform a controlled re-summation of the ratio of exponentials in Eq.~\eqref{eq:conditional_distribution_exp_ratio}.
For an arbitrary site $x \in \Lambda$, let $f_x := (f_{x,U})_{U \ni x}$ be a \textit{family} of real-valued functions $f_{x,U} : \mathcal{S} \rightarrow \mathbb{R}$. 
We require that $f_{x,U}(\s)$ only depends on $\s_U$ (although it need not depend on $s_y$ for every $y \in U$).
For each function in a family $f_x$, we use the supremum norm \begin{align} 
    \norm{f_{x,U}}_\infty := \underset{\s \in \mathcal{S}}{\max} |f_{x,U}(\s)|,
\end{align}
which we use to define a $\mu$-\textit{weighted norm} on the whole family $f_x$:
\begin{align}
    \norm{f_x}_\mu := \sum_{U \ni x} |U| e^{\mu \diam(U)} \norm{f_{x,U}}_\infty.
\end{align}
For any two families $f_x$ and $g_x$, we define their product via \textit{convolution over unions}:
\begin{align}
    (f_x \star g_x)_{x,W}(\s) := \sum_{\substack{U,V \ni x \\ U \cup V = W}} f_{x,U}(\s) \cdot g_{x,V}(\s). \label{eq:convolution_defn}
\end{align}
For any $\s \in \mathcal{S}$, summing over the supports of the full product family gives an ordinary product of sums over the supports of $f_x$ and $g_x$:
\begin{align}
    \sum_{W \ni x} (f_x \star g_x)_{x,W}(\s) = \lrp{\sum_{U \ni x} f_{x,U}(\s)} \cdot \lrp{\sum_{V \ni x} g_{x,V}(\s)}. \label{eq:convolution_sum}
\end{align}
Since all of the supports defining $f_x$ and $g_x$ contain $x$, we have that $\diam(U \cup V) \leq \diam(U) + \diam(V)$ and $|U \cup V| \leq |U| \cdot |V|$.
Therefore, the $\mu$-weighted norm is submultiplicative:
\begin{align}
    \norm{f_x \star g_x}_\mu \leq \norm{f_x}_\mu \cdot \norm{g_x}_\mu.
\end{align}
$I_x=(\ind_{\left\{U = \{ x\}\right\}})_{U \ni x}$ is the identity family, where $\ind_{\{ \cdot \}}$ is the indicator function.
Finally, for a family $f_x$, let $\E_x[f_x] := \E_{s_x \sim \nu_x}[f_x]$ denote the family obtained by averaging each function $f_{x,U}(\s)$ while retaining its unique support $U$. 
\begin{proof}[Proof of Lemma~\ref{lem:stochastic_kernel_decomposition}]
Fix a site $x \in \Lambda$.
The preceding formalism allows us to describe the classical potentials appearing in Eq.~\eqref{eq:conditional_distribution_exp_ratio} as a family of real-valued functions $\Psi_{U} : \mathcal{S} \rightarrow \mathbb{R}$
\begin{align}
    \Psi_x := \lrp{\Psi_{U}}_{U \ni x},
\end{align}
where we omit the explicit $x$-dependence of each function $\Psi_U$ in the family since each classical potential of Eq.~\eqref{eq:conditional_distribution_exp_ratio} only depends on its support $U$.
By assumption, we then have that
\begin{align}
    \norm{\Psi_x}_\mu \leq \mathfrak{g} \leq \log(6/5). \label{eq:Psi_mu_norm_bound}
\end{align}
Exponentiation of the family $\Psi_x$ is defined via the convolution product $\star$:
\begin{align}
    e^{\Psi_x} := \sum_{n \geq 0} \frac{1}{n!} \underbrace{\lrp{\Psi_x \star \cdots \star \Psi_x}}_{\text{$n$ times}} = \sum_{n \geq 0} \frac{1}{n!} (\Psi_x)^{\star n}
\end{align}
Using the assumption in Eq.~\eqref{eq:Psi_mu_norm_bound}, we can bound the $\mu$-norm of the exponential and its average as
\begin{align}
    \norm{e^{\Psi_x} - I_x}_\mu &\leq e^{\mathfrak{g}} - 1, \\
    \norm{\E_x[e^{\Psi_x}] - I_x}_\mu &\leq e^{\mathfrak{g}} - 1 \leq 1/5.
\end{align}
Therefore, the inverse $(\E_x[e^{\Psi_x}])^{-1}$ exists and has bounded $\mu$-norm.
We now define
\begin{align}
    \D_x := \lrp{e^{\Psi_x} - \E_x[e^{\Psi_x}]} \star \lrp{\E_x[e^{\Psi_x}]}^{-1}
\end{align}
which we claim to be the family of functions giving corrections to the single-site distribution $\nu_x$.
By submultiplicativity of the $\mu$-norm and the triangle inequality, we have that 
\begin{align}
    \norm{\D_x}_\mu \leq \frac{2(e^{\mathfrak{g}} - 1)}{2 - e^{\mathfrak{g}}} \leq \frac{1}{2}. \label{eq:D_mu_norm}
\end{align}
Summing over all components of $\D_x$, we use Eq.~\eqref{eq:convolution_sum} to see that
\begin{align}
    \sum_{W \ni x} \D_{x,W}(\s) &= \Big(\sum_{U \ni x}(e^{\Psi_x})_U(\s) - \sum_{U \ni x}(\E_x[e^{\Psi_x}])_U(\s)\Big) \cdot \sum_{V\ni x}\big((\E_x[e^{\Psi_x}])^{-1}\big)_V(\s) \\
    &= \frac{\exp\lrp{\sum_{U \ni x} \Psi_{U}(\s)}}{\E_x\lrb{\exp\lrp{\sum_{V \ni x} \Psi_{V}(\s)}}} - 1
\end{align}
which implies that our conditional distribution can be written in the form
\begin{align}
    \pi(s_x | \s_{x^c}) &= \nu_x(s_x) \lrp{1 + \sum_{U \ni x} \D_{x,U}(\s)}.
\end{align}
For the set of weights $\lambda_{x,U} := \norm{\D_{x,U}}_\infty$, we define a \textit{stochastic kernel} for any $x \in \Lambda$ and any $U \ni x$:
\begin{align}
    \K_{x,U}(s_x | \s_{U \setminus x}) &:= \nu_x(s_x) \lrp{1 + \frac{\D_{x,U}(\s)}{\lambda_{x,U}}}. 
\end{align}
Each kernel is nonnegative since $|\D_{x,U}(\s)| \leq \lambda_{x,U}$ for any $\s \in \mathcal{S}$, and each kernel is normalized since $\E_x[\D_{x,U}] = 0$.
(If $\lambda_{x,U} = 0$, then we just set $\K_{x,U}(s_x|\s_{U \setminus x}) = \nu_x$.)
We can now rewrite the conditional distribution as
\begin{align}
    \pi(s_x|\s_{x^c}) &= \lrp{1 - \sum_{U \ni x} \lambda_{x,U}} \nu_x(s_x) + \sum_{U \ni x} \lambda_{x,U} \K_{x,U}(s_x | \s_{U \setminus x}).
\end{align}
Setting $\lambda_{x,0} := \lrp{1 - \sum_{U \ni x} \lambda_{x,U}}$ gives the exact form for the conditional distribution in Eq.~\eqref{eq:stochastic_kernel_decomposition}, and the $\mu$-norm bound in Eq.~\eqref{eq:D_mu_norm} implies that $\lambda_{x,0} \geq 1/2$.
Furthermore, after pinning the labels of any subset of ${x^c}$, the decomposition remains valid; since the weights are label-independent, we simply fix that subset's labels in the conditioning of each kernel $\K_{x,U}(s_x | \s_{U \setminus x})$.
\end{proof}
\subsubsection{Pairwise influence bound}
For distinct $x,y \in \Lambda$, the \textit{pairwise influence} of site $y$ on site $x$ is defined as
\begin{align}
    \C_{xy} := \sum_{U \ni {x,y}} \lambda_{x,U}, \label{eq:influence_defn}
\end{align}
which sums the weights of all resampling rules (stochastic kernels) for the label $s_x$ that can depend on the label $s_y$.
Using the fact that $\dist(x,y) \leq \diam(U)$ for any $x,y \in U$, the definition of $\lambda_{x,U}$, and the $\mu$-norm bound in Eq.~\eqref{eq:D_mu_norm}, we have that
\begin{align}
    \sum_{y \not= x} \C_{xy} e^{\mu \dist(x,y)} &= \sum_{y \not= x} \sum_{U \ni x,y} \lambda_{x,U} e^{\mu \dist(x,y)} \leq  \sum_{y \not= x} \sum_{U \ni x,y} \lambda_{x,U} e^{\mu \diam(U)} \nonumber \\
    &= \sum_{U \ni x} \sum_{y \in U \setminus x} \lambda_{x,U} e^{\mu \diam(U)} \leq \sum_{U \ni x} |U| e^{\mu \diam(U)} \lambda_{x,U} \nonumber \\
    &= \sum_{U \ni x} |U| e^{\mu \diam(U)} \norm{\D_{x,U}}_\infty = \norm{\D_x}_\mu \leq \frac{1}{2} \label{eq:influence_bound}
\end{align}
which means that the total influence of all sites in $x^c$ on resampling the label $s_x$, weighted by exponentials in their distance to $x$, is at most $1/2$.

\subsection{Local contraction of pairwise CMI}\label{subsec:local_contraction}
We first introduce some notation which will be used in the proof of Lemma~\ref{lem:conditional_information_contraction}. 
Let $\mathcal{X} = \bigotimes_{x \in \Lambda} \mathcal{X}_x$ be an auxiliary Hilbert space, where $\mathcal{X}_x$ is another finite-dimensional Hilbert space for site $x$. 
We define $\mathcal{X}_x$ so that it contains a set of orthogonal pure states $\left\{\ketbra{s_x}\right\}$ corresponding to the labels $s_x \in \mathcal{S}_x$, so $\dim (\mathcal{X}_x) \geq |\mathcal{S}_x|$. 
For our purposes, each $\mathcal{X}_x$ is simply a `classical register' for the label on site $x$.
Define a state $\Omega$ on the enlarged Hilbert space $\mathcal{X} \otimes \mathcal{H}$ as
\begin{align}
    \Omega := \sum_{\s \in \mathcal{S}} \pi(\s) \cdot \ketbra{\s} \otimes \sigma(\s) \label{eq:Omega_defn}
\end{align}
where $\ketbra{\s} := \bigotimes_{x \in \Lambda} \ketbra{s_x}$ is a pure state on $\mathcal{X}$ and $\sigma(\s)$ is a product state\footnote{In our application, we are taking $\sigma(\s) := \bigotimes_{x \in \Lambda} P_{s_x}$. 
While our proof requires that $\sigma(\s)$ is a product state, the state on each site need not be a projector.} on $\mathcal{H}$.
The enlarged state $\Omega$ reduces to the Gibbs state $\rho_\beta$ after tracing out $\mathcal{X}$.
For simplicity of notation in what follows, for any $A \subset \Lambda$ and label set $\s_A \in \mathcal{S}_A$, we will also write $\ketbra{\s_A} := \bigotimes_{x \in A} \ketbra{s_x}$ as a pure label state on $\mathcal{X}_A$ and $\sigma(\s_A)$ as a product state on $\mathcal{H}_A$.

For any qudit region denoted $L \subseteq \Lambda$, let $\ell \in \mathcal{S}_L$ be a fixed label for sites in $L$, and denote the state $\Omega$ with all labels in $L$ pinned to $\ell$ by
\begin{align}
    \Omega^{\ell} = \sum_{\s \in \mathcal{S}}\pi^\ell(\s) \cdot \ketbra{\s} \otimes \sigma(\s) \label{eq:pinned_Omega_defn}
\end{align}
where 
\begin{align}
    \pi^\ell(\s) := \frac{\pi(\s)}{\pi_{L}(\s)} \cdot \ind_{\{\s_L = \ell\}} \label{eq:pinned_distribution}
\end{align}
is the pinned distribution for $\ell$ and $\pi_{L}(\s)$ is the marginal distribution of $\pi(\s)$ on $\mathcal{S}_L$.
Note that, for any site $x \in \Lambda \setminus L$, we can rewrite the state as
\begin{align}
    \Omega^\ell &= \sum_{\s} \pi^\ell(s_x | \s_{x^c}) \pi_{x^c}^\ell(\s_{x^c}) \cdot \ketbra{\s} \otimes \sigma(\s) \nonumber \\
    &= \sum_{\s_{x^c}} \pi_{x^c}^\ell(\s_{x^c}) \cdot \ketbra{\s_{x^c}} \otimes \sigma(\s_{x^c}) \otimes \lrp{\sum_{s_x} \pi^\ell(s_x | \s_{x^c}) \cdot \ketbra{s_x} \otimes \sigma(s_x)}. \label{eq:rewritten_pinned_Omega_1}
\end{align}
Then, under the assumptions of Lemma~\ref{lem:stochastic_kernel_decomposition}, we can use the decomposition in Eq.~\eqref{eq:stochastic_kernel_decomposition} to rewrite the expression in line \eqref{eq:rewritten_pinned_Omega_1} as
\begin{align}
    \Omega^\ell &= \lambda_{x,0} \Omega^{\ell}_0 + \sum_{U \ni x} \lambda_{x,U} \Omega^{\ell}_U \label{eq:pinned_Omega_branch_decomposition}
\end{align}
where we have defined different `branch states' 
\begin{align}
    \Omega^{\ell}_U &:= \sum_{\s_{x^c}} \lrb{\pi_{x^c}^\ell(\s_{x^c})  \ketbra{\s_{x^c}} \otimes \sigma(\s_{x^c}) \otimes \sum_{s_x} \K_{x,U}^{\ell}(s_x | \s_{U \setminus x}) \ketbra{s_x} \otimes \sigma(s_x)} \label{eq:Omega_U_defn}
\end{align}
where, for $\Omega^{\ell}_0$, we replace the kernel $\K_{x,U}^{\ell}(s_x | \s_{U \setminus x})$ with the on-site distribution $\nu_x(s_x)$.
Therefore, the pinned quantum-classical state $\Omega^\ell$ can be written as a convex sum of states whose weights and marginals on $x$ correspond to different resampling branches (labeled by supports $U \ni x$).

Let $A \subseteq \Lambda$ be an arbitrary subset of qudits, and fix an arbitrary site $a \in A$.
For any $x \in \Lambda$ with $x \not= a$, we quantify the maximum pairwise CMI under pinning via
\begin{align}
    \F_{xa} := \underset{\substack{L \subseteq \Lambda \setminus \left\{a,\,x \right\}, \; \ell \in \mathcal{S}_L, \\ B \subseteq \Lambda \setminus \{a\}}}{\sup} \Big[ I(\mathcal{H}_a : \mathcal{X}_x |\mathcal{H}_B)_{\Omega^{\ell}} \Big], \label{eq:pairwise_CMI_defn}
\end{align}
and we let $\F_{aa} = 2\log(q)$.
The supremum in Eq.~\eqref{eq:pairwise_CMI_defn} allows $x \in B$ and runs over every pinning of the labels $\s_L = \ell$ on a subset $L \subseteq \Lambda \setminus \{a,x\}$. 
We have a uniform bound 
\begin{align}
    0 \leq \F_{xa} \leq 2\log(q)
\end{align}
for any $a,x \in \Lambda$ since the CMI is nonnegative with a maximum value of $2\log(\dim\mathcal{H}_a) \leq 2\log(q)$.

We claim that the influence $\C$, defined in Eq.~\eqref{eq:influence_defn}, contracts $\F$, i.e. that
\begin{align}
    \F_{xa} \leq \sum_{y \not= x} \C_{xy} \F_{ya}.\label{eq:pairwise_CMI_contractive}
\end{align}
We defer its proof to Section~\ref{subsec:local_contraction_pairwise_CMI}.
Assuming Eq.~\eqref{eq:pairwise_CMI_contractive} holds, we can straightforwardly prove that
\begin{align}
    \F_{xa} \leq 2 \log(q) e^{-\mu \dist(x,a)}. \label{eq:pairwise_CMI_bound}
\end{align}
\begin{proof}[Proof of Eq.~\eqref{eq:pairwise_CMI_bound}]
Define the site $x'$ as
\begin{align}
    x' := \underset{x \in \Lambda}{\operatorname{argmax}} \lrp{\F_{xa} e^{\mu \dist(x,a)}}. \label{eq:argmax}
\end{align}
Assume for the sake of contradiction that
\begin{align}
    \F_{x'a} e^{\mu \dist(x',a)} > 2\log(q). \label{eq:afsoc}
\end{align}
Since $\F_{aa} e^{\mu \dist(a,a)} = 2\log(q)$, we thus have that $x' \not= a$.
We then find that
\begin{align}
    \F_{x'a} e^{\mu \dist(x',a)} &\leq \sum_{y \not= x'} \C_{x',y} \F_{ya} e^{\mu \dist(x',a)} \tag{by Eq.~\eqref{eq:pairwise_CMI_contractive}} \\
    &\leq \sum_{y\not= x'} \C_{x'y} \F_{ya} e^{\mu \dist(x',y)} e^{\mu \dist(y,a)} \nonumber \tag{by triangle inequality} \\
    &\leq \F_{x'a} e^{\mu \dist(x',a)} \sum_{y \not= x'} \C_{x'y} e^{\mu \dist(x',y)}\tag{by Eq.~\eqref{eq:argmax}} \\
    &\leq \frac{1}{2} \F_{x'a} e^{\mu \dist(x',a)}. \tag{by Eq.~\eqref{eq:influence_bound}}
\end{align}
Since $\F_{x'a} e^{\mu \dist(x',a)}$ is positive by the assumption of Eq.~\eqref{eq:afsoc}, this is a contradiction.
Therefore, the assumption in Eq.~\eqref{eq:afsoc} is false, and we have that
\begin{align}
    \F_{x'a} e^{\mu \dist(x',a)} \leq 2 \log(q). \label{eq:F_exp_distance_maximum}
\end{align}
Since we define $x'$ to maximize the LHS of Eq.~\eqref{eq:F_exp_distance_maximum}, this implies that the bound
\begin{align}
    \F_{xa} e^{\mu \dist(x,a)} \leq 2 \log(q) \nonumber
\end{align}
holds for every $x \in \Lambda$, which completes the proof of Eq.~\eqref{eq:pairwise_CMI_bound}.
\end{proof}

\subsubsection{Influence contracts pairwise CMI}\label{subsec:local_contraction_pairwise_CMI}
We now complete the deferred proof that the influence $\C$ contracts the maximum pairwise CMI $\F$.
\begin{proof}[Proof of Eq.~\eqref{eq:pairwise_CMI_contractive}]
Let $a \in A$ be arbitrary and fixed, and let $x \in \Lambda$ be another site $x \not= a$. 
Let $L \subseteq \Lambda \setminus \{a,x\}$ be an arbitrary subset of qudits, and pin the label set on $L$ to be $\s_L \in \mathcal{S}_L$.
Let $B \subseteq \Lambda \setminus \{a\}$ be another arbitrary subset of qudits.

In the remainder of the proof, we drop explicit tensor product notation for simplicity, instead writing $\mathcal{X}_x \mathcal{H}_x := \mathcal{X}_x \otimes \mathcal{H}_x$.
Let $\Omega^{\ell}$ be the pinned state on $\mathcal{X}\mathcal{H} $.
We first use strong subadditivity to see that
\begin{align}
    I\lrp{\mathcal{H}_a : \mathcal{X}_x \mid \mathcal{H}_B}_{\Omega^{\ell}} \leq I\lrp{\mathcal{H}_a : \mathcal{X}_x \mathcal{H}_x \mid \mathcal{H}_{B\setminus x}}_{\Omega^{\ell}} \label{eq:CMI_contraction_step1}
\end{align}
which holds both when $x \in B$ and when $x \not\in B$.

Our weights satisfy $\lambda_{x,0} + \sum_{U \ni x} \lambda_{x,U} = 1$ and each is nonnegative.
Furthermore, $\Omega^\ell_0$ and the set of $\{\Omega^\ell_U\}_{U \ni x}$ all have the same marginal on $\mathcal{H}_a \mathcal{H}_{B \setminus \{ x \}}$. 
Then, by applying the concavity of the conditional entropy to the pinned state in Eq.~\eqref{eq:pinned_Omega_branch_decomposition}, we have
\begin{align}
    I\lrp{\mathcal{H}_a : \mathcal{X}_x \mathcal{H}_x \mid \mathcal{H}_{B\setminus x}}_{\Omega^{\ell}} \leq \sum_{U \ni x} \lambda_{x,U} I\lrp{\mathcal{H}_a : \mathcal{X}_x \mathcal{H}_x \mid \mathcal{H}_{B\setminus x}}_{\Omega^{\ell}_U} \label{eq:CMI_contraction_step2}
\end{align}
where the CMI contribution from the state $\Omega^\ell_0$ vanishes because this corresponds to a resampling branch in which $s_x$ is independently drawn from $\nu_x$.
Now, let $U \ni x$ be the support of a resampling branch, and let $Y = U \setminus (L \cup \{x\})$ be the set of unpinned sites in $U$ that it conditions on.
Then, consider a channel $\mathcal{M}_U^\ell : \mathcal{X}_Y \rightarrow \mathcal{X}_x \mathcal{H}_x $ which corresponds to
\begin{enumerate}[(i)]
    \item reading the classical labels $\s_Y$ on sites in $Y$,
    \item sampling a label $s_x$ at site $x$ according to $\K_{x,U}^{\ell}\lrp{s_x|\s_Y}$, and then
    \item retaining the classical label $s_x$ and preparing the corresponding state $\sigma(s_x)$ on $\mathcal{H}_x$.
\end{enumerate}
This is a valid quantum channel, and it does not act on $\mathcal{H}_a$ or on $\mathcal{H}_{B \setminus x}$.
Therefore, by the data processing inequality, we have for each $U \ni x$ that
\begin{align}
    I\lrp{\mathcal{H}_a : \mathcal{X}_x \mathcal{H}_x \mid \mathcal{H}_{B\setminus x}}_{\Omega^{\ell}_U} \leq I\lrp{\mathcal{H}_a : \mathcal{X}_{Y} \mid \mathcal{H}_{B\setminus x}}_{\Omega^{\ell}}. \label{eq:CMI_contraction_step3}
\end{align}

To further bound the RHS of Eq.~\eqref{eq:CMI_contraction_step3}, we use the chain rule for CMI.
Indexing the sites in $Y = \left\{y_1,\ldots,y_{|Y|}\right\}$, define the truncated set $Y_{(k)} := \left\{y_1,\ldots,y_{k}\right\}$ for $1 \leq k < |Y|$ (and otherwise, $Y_{(k)} = \varnothing$).
If $a \in Y$, then set $y_{|Y|} = a$; otherwise, the ordering of sites in $Y$ is arbitrary.
Then, the chain rule for CMI gives
\begin{align}
    I\lrp{\mathcal{H}_a : \mathcal{X}_{Y} \mid \mathcal{H}_{B\setminus x}}_{\Omega^{\ell}} = \sum_{1 \leq k \leq |Y|} I(\mathcal{H}_a : \mathcal{X}_{y_k} \mid \mathcal{X}_{Y_{(k-1)}} \mathcal{H}_{B\setminus x})_{\Omega^\ell}. \label{eq:CMI_contraction_step4}
\end{align}
Note that the CMI expressions in the sum condition on the `classical registers' for the labels, and thus the additional conditioning on $\mathcal{X}_{Y_{(k-1)}}$ corresponds to pinning the labels on some sites in $Y$. 
Each CMI in the sum can be expressed as a convex combination of terms, each bounded by
\begin{align}
    I(\mathcal{H}_a : \mathcal{X}_{y_k} \mid \mathcal{H}_{B \setminus x})_{\Omega^{\ell'}} \leq \F_{y_k a} \label{eq:chain_rule_term_bound}
\end{align}
where $\ell' = (\ell,\s_{Y_{(k-1)}})$ for some label assignment $\s_{Y_{(k-1)}}$.
In the case that $a \in Y$ and thus $y_{|Y|} = a$, the bound still holds since $\F_{aa} = 2\log(q)$.
Putting everything together, we have the following chain of inequalities:
\begin{align}
    I\lrp{\mathcal{H}_a : \mathcal{X}_x \mid \mathcal{H}_B}_{\Omega^{\ell}} &\leq I\lrp{\mathcal{H}_a : \mathcal{X}_x \mathcal{H}_x \mid \mathcal{H}_{B\setminus x}}_{\Omega^{\ell}} \tag{by Eq.~\eqref{eq:CMI_contraction_step1}} \\
    &\leq \sum_{U \ni x} \lambda_{x,U} I\lrp{\mathcal{H}_a : \mathcal{X}_x \mathcal{H}_x \mid \mathcal{H}_{B\setminus x}}_{\Omega^{\ell}_U} \tag{by Eq.~\eqref{eq:CMI_contraction_step2}} \\
    &\leq \sum_{U \ni x} \lambda_{x,U} I(\mathcal{H}_a : \mathcal{X}_{U \setminus (L \,\cup\, x)} \mid \mathcal{H}_{B \setminus x})_{\Omega^\ell} \tag{by Eq.~\eqref{eq:CMI_contraction_step3}} \\
    &\leq \sum_{U \ni x} \lambda_{x,U} \sum_{y \in U \setminus (L \,\cup\, x)}  \F_{ya} \tag{by Eqs. \eqref{eq:CMI_contraction_step4} and \eqref{eq:chain_rule_term_bound}} \\
    &\leq \sum_{U \ni x} \lambda_{x,U} \sum_{\substack{y \in U, \\ y \not= x}}  \F_{ya} \\
    &= \sum_{y \not= x} \lrp{\sum_{U \ni x,y} \lambda_{x,U}} \F_{ya} = \sum_{y \not= x} \C_{xy} \F_{ya}. \label{eq:final_CMI_contraction_inequality}
\end{align}
Our choices of $a$, $x$, $B$, $L$, and the pinned labels $\ell$ were arbitrary, so the CMI upper bound in line~\eqref{eq:final_CMI_contraction_inequality} holds for $\F_{xa}$. 
This proves the claim.
\end{proof}
\subsubsection{CMI as a sum of decaying exponentials}
\begin{proof}[Proof of Lemma~\ref{lem:conditional_information_contraction}]
Let $a, c \in \Lambda$ be distinct sites.
Then, for any $B'$ not containing $a$ or $c$, the state $\Omega_{\mathcal{H}_a \mathcal{H}_c \mathcal{H}_{B'}}$ can be prepared from $\Omega_{\mathcal{H}_a \mathcal{X}_c \mathcal{H}_{B'}}$ via a channel which reads the label $s_c$ and prepares the state $\sigma(s_c)$ on $\mathcal{H}_c$. 
Using the data-processing inequality, Eq.~\eqref{eq:pairwise_CMI_defn}, and Eq.~\eqref{eq:pairwise_CMI_bound} respectively, we obtain
\begin{align}
    I(\mathcal{H}_a : \mathcal{H}_c \mid \mathcal{H}_{B'})_{\rho_\beta} &\leq I(\mathcal{H}_a : \mathcal{X}_c \mid \mathcal{H}_{B'})_{\Omega} \leq \F_{ca} \leq 2\log(q) e^{-\mu \dist(a,c)}. \label{eq:pairwise_CMI_decay}
\end{align}

Finally, let $A \sqcup B \sqcup C \subseteq \Lambda$ be an arbitrary tripartition, and order the sites as $A = \left\{a_1,\ldots,a_{|A|}\right\}$ and $C = \left\{c_1,\ldots,c_{|C|}\right\}$.
Applying the chain rule for CMI to both $\mathcal{H}_A$ and $\mathcal{H}_C$, we have
\begin{align}
    I(\mathcal{H}_A : \mathcal{H}_C \mid \mathcal{H}_B)_{\rho_\beta} &= \sum_{i=1}^{|A|} \sum_{j=1}^{|C|} I(\mathcal{H}_{a_i} : \mathcal{H}_{c_j} \mid \mathcal{H}_B \mathcal{H}_{a_1}\cdots\mathcal{H}_{a_{i-1}} \mathcal{H}_{c_1}\cdots \mathcal{H}_{c_{j-1}}) \\
    &\leq 2 \log(q) \sum_{a\in A, \; c \in C} e^{-\mu \dist(a,c)} \label{eq:sum_of_exponentials_CMI_bound}
\end{align}
where the inequality is obtained by applying Eq.~\eqref{eq:pairwise_CMI_decay} to each CMI term from the chain rule.
This completes the proof.
\end{proof}

\subsection{Exponential decay of conditional mutual information}\label{subsec:CMI_proof_completion}
Our final lemma, which is needed to convert the bound from Lemma~\ref{lem:conditional_information_contraction} into the bound of Theorem~\ref{thm:CMI_decay}, separates a shortest path into an initial segment inside $A$, a segment between the two boundaries, and a final segment inside $C$.
The sums over the first and last segments cost only structural constants.

\begin{lemma}[Boundary summation]\label{lem:geometry}
Let $G=(V,E)$ be a finite graph with shortest-path distance $d_G$, and let $\alpha>0$.
Distances between distinct connected components are $+\infty$, with $e^{-\alpha\infty}=0$. 
For $A\subseteq V$, define
\begin{align}\label{eq:graph-boundary-kernel}
 \partial_G A=\{a\in A:\exists x\in V\setminus A,\ \{a,x\}\in E\},
 \qquad
 L_G(\alpha)=\sup_{u\in V}\sum_{x\in V}e^{-\alpha d_G(u,x)}.
\end{align}
Then, for disjoint $A,C\subseteq V$ and $\alpha>0$, we have
\begin{align}\label{eq:geometry}
 \sum_{a\in A,c\in C}e^{-\alpha d_G(a,c)}
 \le L_G(\alpha)^2
     \sum_{u\in\partial_G A,v\in\partial_G C}
             e^{-\alpha d_G(u,v)}.
\end{align}
\end{lemma}

\begin{proof}
For each pair $(a,c)$ at finite distance, choose a shortest path from $a$
to $c$. Let $u$ be its vertex immediately before its first exit from $A$,
and let $v$ be its vertex at its last entry into $C$. Then
$u\in\partial_G A$, $v\in\partial_G C$, and $u$ precedes $v$. Every segment
of a shortest path is shortest, hence
\begin{align}\label{eq:geodesic-decomposition}
 d_G(a,c)=d_G(a,u)+d_G(u,v)+d_G(v,c).
\end{align}
Assign $(a,c)$ to $(u,v)$. Dropping the assignment restrictions, the contribution assigned to fixed $u,v$ is at most
\begin{align}\label{eq:assigned-pair-bound}
 e^{-\alpha d_G(u,v)}
 \left(\sum_{a\in A}e^{-\alpha d_G(a,u)}\right)
 \left(\sum_{c\in C}e^{-\alpha d_G(v,c)}\right)
 \le L_G(\alpha)^2e^{-\alpha d_G(u,v)}.
\end{align}
Summing proves Eq.~\eqref{eq:geometry}; infinite-distance pairs contribute zero.
The same proof applies to infinite graphs when the endpoint sets are finite and $L_G(\alpha)<\infty$.
\end{proof}

With Lemma~\ref{lem:geometry} in hand, we now complete the proof of Theorem~\ref{thm:CMI_decay}. 

\begin{proof}[Proof of Theorem~\ref{thm:CMI_decay}]
Assume $q, k \geq 2$ and $J>0$, since otherwise the decay of CMI is immediate.
First, we check the stronger weight needed in Lemma~\ref{lem:expansion}.
Writing $\varepsilon=\beta J$, the threshold in Eq.~\eqref{eq:CMI_decay_threshold_thm} gives $\varepsilon<1$ and
\begin{align}\label{eq:cmi-temperature}
 \eta_{k+2}=\varepsilon(2q-1)^k e^{\varepsilon+k+3}
 \le\frac1{8k}\left(\frac{2q-1}{2q}\right)^k e^{\varepsilon+1-k}
 \le\frac1{8k}.
\end{align}
Thus, the scalar expansion gives the weighted bound in Eq.~\eqref{eq:potential} for the classical interaction.

The interaction graph $G_H$ has vertex set $\Lambda$ and joins distinct $x,y \in \Lambda$ whenever both sites belong to some $X\in\mathcal E$.
Its inner graph boundary is given by Eq.~\eqref{eq:interaction_boundary_defn}.
Every interaction support is a clique, so it has graph diameter at most one.

Every interaction lies in one connected component of $G_H$.
Consequently the Hamiltonian splits across components, its Gibbs state is their tensor product, and CMI of its marginals is additive across components.
It suffices to apply the following arguments within each component; cross-component pairs have zero exponential weight.

Apply Lemma~\ref{lem:expansion} with metric $d_H:=d_{G_H}$ and range $r=1$.
Since $|U|\le e^{|U|}$, Eq.~\eqref{eq:potential} gives
\begin{align}\label{eq:interaction-graph-potential}
 \sup_x\sum_{U\ni x}|U|e^{2\diam_{d_H}(U)}
                 \norm{\Psi_U}_\infty
 \le2\eta_{k+2}\le\frac1{4k}\le\frac18<\log(6/5).
\end{align}
Lemma~\ref{lem:conditional_information_contraction} with decay exponent $\mu = 2$ therefore implies
\begin{align}\label{eq:pairwise}
 I(A:C\mid B)_{\rho_{\beta,ABC}}
 \le2\log(q)\sum_{a\in A,c\in C}e^{-2d_H(a,c)}.
\end{align}
The preparation identity commutes with the partial trace:
\begin{align}\label{eq:reduced-preparation}
 \rho_{\beta,ABC}=\underset{\s \sim \pi}{\E} \left[ \bigotimes_{x\in A \sqcup B \sqcup C}P_{s_x} \right].
\end{align}
All hidden labels, including those in $D$, are averaged over.

Each graph edge joins sites at lattice distance at most $R$. Hence
\begin{align}\label{eq:comparison}
 d_H(x,y)\ge d(x,y)/R,
 \qquad L_{G_H}(\alpha)\leq \mathsf{S}_{R/\alpha}.
\end{align}
Applying Lemma~\ref{lem:geometry} to Eq.~\eqref{eq:pairwise} with $\alpha=2$ gives
\begin{align}\label{eq:cmi-boundary-sum}
 I(A:C\mid B)_{\rho_{\beta,ABC}}
 \le2\log(q) \,\mathsf{S}_{R/2}^{\,2}
       \sum_{u\in\partial A,v\in\partial C}e^{-2d_H(u,v)}.
\end{align}
If there is no finite-distance endpoint pair, the CMI vanishes.
Otherwise, set $\ell=d_H(A,C)$. Since $d_H(u,v)\ge\ell$ on the boundary sum, summing from either endpoint yields
\begin{align}\label{eq:cmi-boundary-decay}
 \sum_{u\in\partial A,v\in\partial C}e^{-2d_H(u,v)}
 &\le e^{-\ell}
       \sum_{u\in\partial A,v\in\partial C}e^{-d_H(u,v)}
\\
 &\le e^{-\ell} \, \mathsf{S}_R
       \min\{|\partial A|,|\partial C|\}
\\
 &\le e^{-\dist(A,C)/R} \, \mathsf{S}_R
       \min\{|\partial A|,|\partial C|\}.
\end{align}
This completes the proof.
\end{proof}

\section{Discussion}\label{sec:discussion}
\subsection*{High-temperature Gibbs states are trivial mixed-state phases of matter} \label{subsec:review_mixed_state_phases}
A widely used definition of mixed-state phases is two-way connectivity: $\rho$ and $\sigma$ lie in the same phase if one can be approximately mapped to the other by a finite-depth local channel (or by a short-time quasi-local Lindbladian evolution)~\cite{Coser_2019,Sang2024,sang2023mixedstate,Ma2023}. 
Since channels are irreversible, such a definition can identify states with distinct topological properties to be in the same phase. 
This motivates the refined requirement that the connecting channels be locally reversible, which is guaranteed when the Markov length remains finite along the evolution \cite{Sang2024,sang2025mixedstatephaseslocalreversibility}.
Such locally-reversible circuits have been constructed for quantum Gibbs states satisfying a ``stable clustering'' assumption~\cite{Ma2025Gibbs}. 

\begin{figure}[!htbp]
\centering
\begin{tikzpicture}[x=1cm,y=1cm,font=\small,line width=0.8pt]
  \definecolor{coldbulk}{RGB}{62,108,153}
  \definecolor{hotbulk}{RGB}{221,145,72}
  \filldraw[fill=coldbulk!12,draw=coldbulk!75!black]
    (0,0) rectangle (3.5,5.7);
  \filldraw[fill=hotbulk!20,draw=hotbulk!80!black]
    (3.5,0) rectangle (7,5.7);
  \draw[black!70,line width=1.1pt] (3.5,0) -- (3.5,5.7);
  \node[align=center,text width=2.9cm,text=coldbulk!65!black] at (1.75,4.85)
    {Finite-$\beta$\\Gibbs state\\[5pt]$\rho_L\propto e^{-\beta H_L}$};
  \node[align=center,text width=2.9cm,text=hotbulk!60!black] at (5.25,4.85)
    {Maximally mixed\\state ($\beta_R=0$)\\[5pt]$\rho_R\propto \id$};
  \node[font=\footnotesize,align=center,text width=6.3cm] at (3.5,-0.3)
    {Both Gibbs states lie in the low-$\beta$ regime};
  \draw[black!80,line width=1pt] (3.5,2.85) circle (0.36);
  \draw[gray!65,dashed] (3.76,3.10) -- (7.5,5.0);
  \draw[gray!65,dashed] (3.76,2.60) -- (7.5,0.7);
  \node[anchor=west,align=left] at (10.25,4.78)
    {$I(C\,{:}\,E| B)$\\[4pt]
     $\displaystyle\lesssim |\partial C|e^{-\operatorname{dist}(C,E)/R}$};
  \node[anchor=west,align=left] at (10.25,2.88)
    {$I(A\,{:}\,C)$\\[4pt]
     $\displaystyle\lesssim |\partial C|e^{-\operatorname{dist}(A,C)/R}$};
  \node[anchor=west,align=left] at (10.25,0.98)
    {$I(A\,{:}\, C| B)$\\[4pt]
     $\displaystyle\lesssim |\partial C|e^{-\operatorname{dist}(A,C)/R}$\\[4pt]
     {\footnotesize $D$ traced out}};
  \begin{scope}[shift={(8.85,4.78)}]
    \fill[cyan!18,rounded corners=4pt] (-1.16,-0.82) rectangle (1.16,0.82);
    \filldraw[fill=cyan!9] (0.12,0) circle (0.65);
    \filldraw[fill=white] (0.12,0) circle (0.32);
    \node at (0.12,0) {$C$};
    \node at (-0.36,0) {$B$};
    \node at (0.99,0) {$E$};
    \node[anchor=south west,font=\small\bfseries] at (-1.2,-0.77) {P0};
    \foreach \ang in {45,90,135,225,270,315}
      \draw[->,cyan!60!black,thin] (\ang:0.57) ++(0.12,0) -- ++(\ang:-0.19);
  \end{scope}
  \begin{scope}[shift={(8.85,2.88)}]
    \fill[green!15,rounded corners=12pt] (-1.16,-0.82) rectangle (1.16,0.82);
    \filldraw[fill=green!6] (0.12,0) circle (0.65);
    \filldraw[fill=green!30] (0.12,0) circle (0.32);
    \node at (0.12,0) {$C$};
    \node at (-0.82,0.12) {$A$};
    \node[anchor=south west,font=\small\bfseries] at (-1.20,-0.77) {M0};
  \end{scope}
  \begin{scope}[shift={(8.85,0.98)}]
    \fill[green!15,rounded corners=12pt] (-1.16,-0.82) rectangle (1.16,0.82);
    \filldraw[fill=green!6] (0.12,0) circle (0.65);
    \fill[green!40] (0.12,0.65) arc[start angle=90,end angle=270,radius=0.65] -- cycle;
    \draw (0.12,0) circle (0.65);
    \draw (0.12,-0.65) -- (0.12,0.65);
    \filldraw[fill=green!30] (0.12,0) circle (0.32);
    \node at (0.12,0) {$C$};
    \node at (-0.37,0) {$B$};
    \node at (0.61,0) {$D$};
    \node at (-0.82,0.12) {$A$};
    \node[anchor=south west,font=\small\bfseries] at (-1.20,-0.77) {M1};
  \end{scope}
\end{tikzpicture}
\caption{{\bf Stitching high-temperature Gibbs states for the mixed-state bootstrap.}
A high-temperature Gibbs state at inverse temperature $\beta$ (left) is joined to a maximally mixed state (right).
The right bulk has zero Hamiltonian, and interactions crossing the interface are zero.
The conditions \textbf{P0}, \textbf{M0}, and \textbf{M1} hold in both bulks and across the interface, up to errors that vanish under coarse-graining.}
\label{fig:high-temperature-interface}
\end{figure}
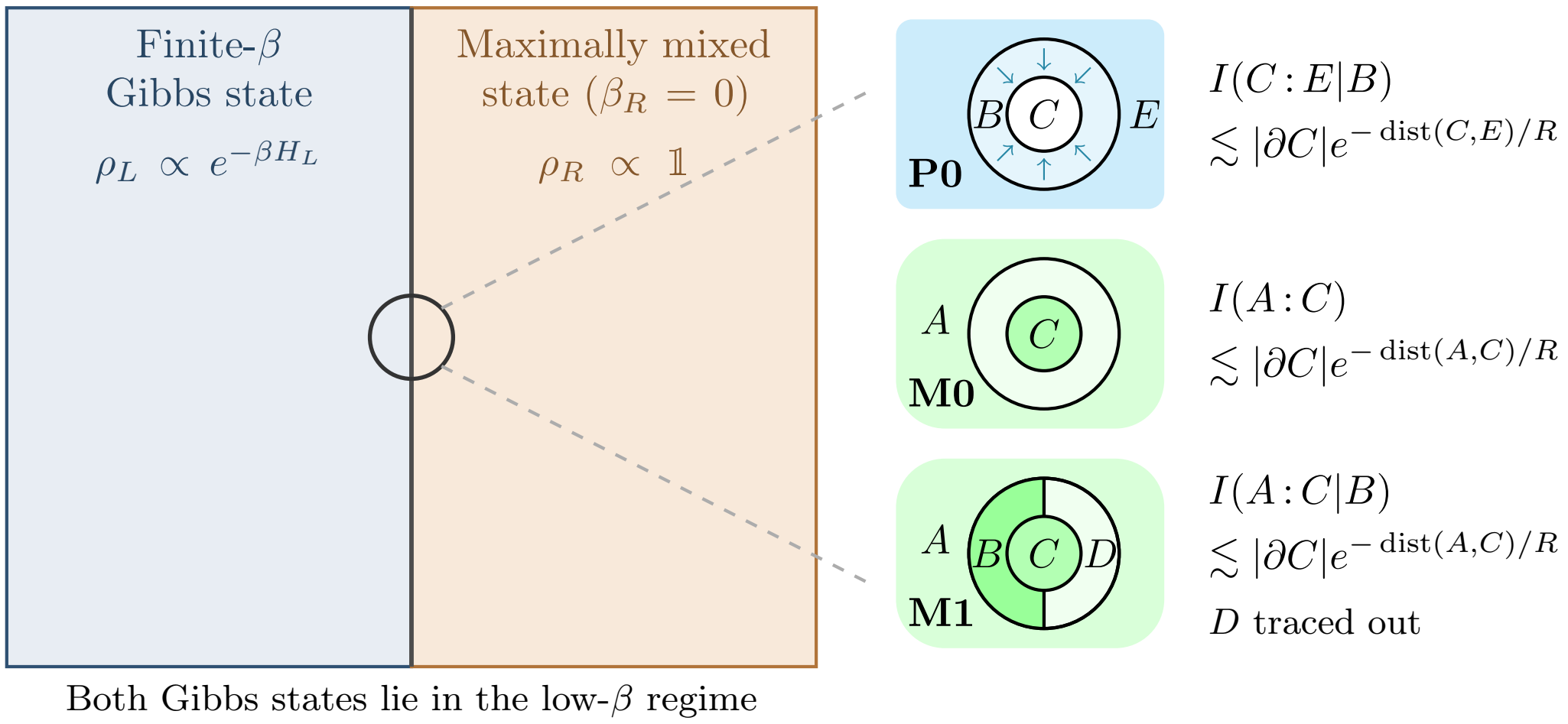

The mixed-state entanglement bootstrap gives a complementary approach to defining mixed-state phases~\cite{yang2025topologicalmixedstatesphases}.
In this framework, one characterizes the fixed-point state of a phase by three structural properties, illustrated in Fig.~\ref{fig:high-temperature-interface}: 
\textbf{P0}, vanishing CMI for an annulus shielding a disk from the rest of the system (local recoverability); 
\textbf{M0}, vanishing MI between distant regions (no long-range order); 
and \textbf{M1}, vanishing CMI for a non-shielding partition of a ball with part of it traced out (space uniformity). 
Then, by relaxing these three structural conditions from exact vanishing to exponential decay in the distance between their arguments, we reach a state away from the fixed point.
Two states lie in the same phase if they can be joined by a domain wall across which all three properties hold, up to errors that vanish under coarse-graining. 

Our results establish that Gibbs states at sufficiently high temperature belong to the trivial phase under both criteria.
First, since Theorem~\ref{thm:MI_decay}  applies uniformly throughout a sufficiently small neighborhood of local Hamiltonian coefficients within the high-temperature regime, its MI bound implies stable clustering via Pinsker's inequality.
Moreover, our CMI bound upgrades the construction of locally-reversible circuits~\cite{Ma2025Gibbs}, which connects $\rho_\beta$ to the maximally mixed state along the path $\rho_s\propto e^{-s\beta H}$, $0\leq s\leq1$, from polylogarithmic to constant depth, with gate diameter from $\mathcal O(\log^D(n/\epsilon))$ to $\mathcal{O}(\log(n/\epsilon))$ for global trace-norm error $\epsilon$.

For the bootstrap criterion, we stitch two high-temperature Gibbs states across an interface, as illustrated in Fig.~\ref{fig:high-temperature-interface} as $\rho = \rho_L \otimes \rho_R$.
Writing their inverse temperatures as $0\leq\beta_R\leq\beta_L$, the hotter state $e^{-\beta_R H_R}/Z_R$ is equivalently a Gibbs state at inverse temperature $\beta_L$ with weaker Hamiltonian $(\beta_R/\beta_L)H_R$.
The total system is therefore a Gibbs state at a common temperature with spatially varying interactions.
Theorems~\ref{thm:MI_decay} and~\ref{thm:CMI_decay} can then be applied across the interface, so all three bootstrap conditions hold up to errors that vanish under coarse-graining.
This establishes phase equivalence between two Gibbs states. Taking $\beta_R=0$ identifies the phase with the maximally mixed state.

\subsection*{Stability of high-temperature Gibbs state phases} \label{subsec:stability}
A natural question is whether the exponential decay of MI and CMI survives local noise~\cite{ZhangGopalakrishnan2026HiddenMarkov,NewsomLevin2026Instability,NegariZouHsieh2026Instability,yang2026stabilityquantumldpcstates}.
Here we consider the output $\widetilde\rho=\mathcal N(\rho_\beta)$ of a local quantum channel and ask whether it continues to satisfy boundary-dependent exponential decay bounds for axioms \textbf{P0}, \textbf{M0}, and \textbf{M1}.

\subsubsection*{Independent on-site channels.}
For $\mathcal N=\bigotimes_{x\in\Lambda}\mathcal N_x$, with arbitrary on-site CPTP maps, MI decay follows directly from data processing:
\begin{equation}
 I(A:C)_{\widetilde\rho}\le I(A:C)_{\rho_\beta}.
\end{equation}
Thus, the bound of Theorem~\ref{thm:MI_decay} persists at the same temperature threshold.
For CMI, stability follows from our classical representation:
\begin{equation}
 \widetilde\rho
 =\sum_{\s}\pi(\s)\bigotimes_{x\in\Lambda}\sigma_x(s_x),
 \qquad \sigma_x(s_x)=\mathcal N_x(P_{s_x}).
\end{equation}
The classical distribution $\pi$ and its effective potentials are unchanged. Only the local density matrices prepared from the labels change.
The CMI contraction argument already allows arbitrary local density matrices $\sigma_x(s_x)$, so the bound of Theorem~\ref{thm:CMI_decay} also persists at its temperature threshold.

\subsubsection*{Weak local channels: conjecture and sketch.}
More generally, let $\mathcal N=\mathcal N^{(m)}\circ\cdots\circ\mathcal N^{(1)}$ be a circuit of fixed depth $m$, where the channel $\mathcal{N}^{(l)}$ at layer $l$ is a tensor product of local quantum channels on disjoint supports. 
We index each of those local channels by $g$, denoting them by $\mathcal N^{(l)}_g$ and their support by $Y^{(l)}_g\subseteq\Lambda$.
Assume that $|Y^{(l)}_g|\le s$, $\diam(Y^{(l)}_g)\le r_0$, and
\begin{align}
 \frac12\|\mathcal N^{(l)}_g-\mathrm{id}_{Y^{(l)}_g}\|_\diamond\le\epsilon,
\end{align}
where $m,s,r_0$ are independent of system size and for a linear map $\Phi$ on operators on $\mathcal H_{Y^{(l)}_g}$, the diamond norm is
\begin{align}
 \|\Phi\|_\diamond
 :=\sup_{X\ne0}
 \frac{\|(\Phi\otimes\mathrm{id}_{E})(X)\|_1}{\|X\|_1},
 \qquad \|X\|_1:=\Tr\sqrt{X^\dagger X},
\end{align}
where $X$ ranges over operators on $\mathcal H_{Y^{(l)}_g}\otimes\mathcal H_E$.

We conjecture that, at sufficiently high temperature and for sufficiently small $\epsilon$, the output state $\tilde{\rho}$ would satisfy
\begin{equation}
 \begin{aligned}
 I(A:C)_{\widetilde\rho}, \, I(A:C| B)_{\widetilde\rho}
 \le K\min\{|\partial^* A|,|\partial^* C|\}
 e^{-\dist(A,C)/\xi}
 \end{aligned}
\end{equation}
for arbitrary disjoint $A,B,C$, with $K,\xi$ independent of system size.
Here, the starred boundary $\partial^* S$ consists of the sites in $S$ touched by a Hamiltonian interaction or a nonidentity channel gate that also touches at least one site outside $S$.
Related interaction-boundary prefactors appear in the bounds on exponential CMI decay for qLDPC states under weak $k$-local Pauli noise in Ref.~\cite{yang2026stabilityquantumldpcstates}.


Writing $\mathcal N^{(l)}_g=\mathrm{id}_{Y^{(l)}_g}+\Delta^{(l)}_g$ suggests a joint cluster expansion of the channel perturbations and Hamiltonian terms in our local $2$-design representation for small $\beta J$ and $\epsilon$.
If one can prove the positivity of the scalar weights and an exponentially weighted bound on their effective potentials preserving strict contraction, then the cluster cancellation argument of Section~\ref{sec:MI_proof} and the information-contraction argument of Section~\ref{sec:CMI_proof} would give the conjectured MI and CMI bounds and thus stability of the three bootstrap conditions.

\subsection*{Bosonic systems and entropic order}
While Theorems~\ref{thm:MI_decay} and~\ref{thm:CMI_decay} show that any local quantum system with bounded local dimension and interaction strength $J$ becomes trivial at sufficiently high temperature, this conclusion can fail in bosonic systems with unbounded local dimension.
In particular, Han et al.~\cite{HanEtAl2025EntropicOrder} recently demonstrated that local bosonic models can exhibit \emph{entropic order} at arbitrarily high temperatures, because ordering in one sector unlocks large fluctuations in auxiliary bosonic modes~\cite{HuangEtAl2025MinimalEntropicOrder,AndrioloEtAl2026ProofEntropicOrder,HsinKobayashi2026EntropicOrders,ZaneLucas2026EntropicSuperfluid}.
In our bounds, the inverse-temperature thresholds scale as $\beta J \lesssim (2q)^{-k}$, which vanish as the local dimension cutoff $q\to\infty$ and the local operator norm $J$ diverges.
Thus, entropic order shows that bounded local dimension and interaction strength are physically essential assumptions for universal high-temperature triviality of the phase of matter, and that classifying phases of systems with infinite local dimension requires further investigation.

\section*{Acknowledgement}

AK is supported by the IIDAI-IBM Accelerator Institute. THY is supported by Taiwan-UIUC Scholarship. JYL acknowledges the support of the Microsoft Quantum Pioneers Program. 
The proof presented in this work was found using GPT-6 Astra under the guidance of the authors.
The authors independently verified and organized the proof, and subsequently improved its exposition for human readability.
GPT-6 Astra and Claude Opus 5.5 were also used to assist with literature search, technical refinements, writing, and editing. 
The authors take full responsibility for the correctness and presentation of the content in the manuscript.

\appendix
\section{The lattice kernel on \texorpdfstring{$\mathbb Z^D$}{Z\textasciicircum D}}\label{app:lattice_kernel}

For nearest-neighbor distance, $d(x,y)=\sum_{j=1}^D|x_j-y_j|$.
Translation invariance makes the sum in Eq.~\eqref{eq:lattice_kernel}
independent of $x$, and the nonnegative summands factorize over coordinates:
\begin{equation}
\mathsf{S}_t=\sum_{z\in\mathbb Z^D}\prod_{j=1}^D e^{-|z_j|/t}
 =\left(\sum_{n\in\mathbb Z}e^{-|n|/t}\right)^D.
\end{equation}
For $t>0$, the one-dimensional sum is a convergent geometric series,
\begin{equation}
 \sum_{n\in\mathbb Z}e^{-|n|/t}
 =1+2\sum_{n=1}^{\infty}e^{-n/t}
 =\frac{1+e^{-1/t}}{1-e^{-1/t}}
 =\coth\!\left(\frac1{2t}\right),
\end{equation}
which implies that $\mathsf{S}_t$ on the $D$-dimensional lattice $\mathbb Z^D$ with nearest-neighbor distance is given as
\begin{equation}
 \mathsf{S}_t=\left[\coth\!\left(\frac1{2t}\right)\right]^D. \label{eq:ZD_lattice_kernel_coth}
\end{equation}

\section{Details of the rooted-tree estimate}\label{app:rooted_tree_count}

We prove Eq.~\eqref{eq:rooted_tree_count}. 
Here, $m=|\gamma|$ is the number of distinct interaction labels in a polymer, not the number of sites in its support. 
The weights $v_X$ are nonnegative, $|X| \leq k$, and Eq.~\eqref{eq:weighted_interaction_sum} gives
\begin{align}
 \sup_y\sum_{X\ni y}v_X\le\eta_\lambda. \label{eq:appendix_site_weight}
\end{align}

\paragraph{Tree representation and counting.}
Each polymer $\gamma$ with $|\gamma|=m$ and $x\in U_\gamma$ admits a spanning tree of its overlap graph, rooted at an interaction containing $x$. 
Ordering the children gives a rooted ordered tree whose vertex labels determine $\gamma$. 
Summing over all such trees therefore overcounts polymers; allowing repeated labels while requiring only parent--child overlap further enlarges the sum.
A depth-first traversal encodes each tree shape uniquely by recording $U$ when moving to a child and $D$ when returning to its parent.
Each of the $m-1$ edges is traversed in both directions, so there are at most $2^{2(m-1)}=4^{m-1}$ shapes.

\paragraph{Summing labels from the leaves to the root.}
For any fixed parent label $X$, the sum over a child label $Y$ obeys
\begin{align}
 \sum_{Y:\,Y\cap X\ne\varnothing}v_Y
 \le\sum_{y\in X}\sum_{Y\ni y}v_Y
 \le |X|\eta_\lambda\le k\eta_\lambda. \label{eq:appendix-child-weight}
\end{align}
The allowed child labels depend on the parent, but this upper bound is uniform in the parent. 
For example, a three-vertex chain gives
\begin{align}
 \sum_{X \ni x} v_X
   \sum_{Y:\,Y\cap X \not= \varnothing} v_Y
   \sum_{Z:\,Z\cap Y \not= \varnothing} v_Z
 &\leq k\eta_\lambda\sum_{X\ni x} v_X
                  \sum_{Y:\,Y\cap X \not= \varnothing} v_Y \\
 &\leq (k\eta_\lambda)^2\sum_{X\ni x} v_X
 \leq \eta_\lambda(k\eta_\lambda)^2.
\end{align}
For an arbitrary fixed tree shape, sum the label of a leaf first.
It appears only in its own weight and in the constraint that it overlap its parent, so \eqref{eq:appendix-child-weight} removes that leaf at a cost of at most $k\eta_\lambda$.
Repeating this procedure removes all $m-1$ nonroot vertices.
The remaining root sum is $\sum_{X\ni x}v_X\le\eta_\lambda$ by Eq.~\eqref{eq:appendix_site_weight}. 
Thus, each tree shape contributes at most $\eta_\lambda(k\eta_\lambda)^{m-1}$. 

Combining this estimate with the number of tree shapes yields
\begin{align}
 \sum_{\substack{\gamma:\,x\in U_\gamma\\|\gamma|=m}}
 \prod_{X\in\gamma}v_X
 \le4^{m-1}\eta_\lambda(k\eta_\lambda)^{m-1}
 =\eta_\lambda(4k\eta_\lambda)^{m-1},
\end{align}
including $m=1$, when only the root remains.

\bibliographystyle{unsrtnat}
\bibliography{bibilography}

\end{document}